\documentclass{article}

\usepackage{arxiv}
\makeatletter
\renewcommand{\LARGE}{\@setfontsize\LARGE\@xivpt{16}}
\makeatother

\usepackage[scaled=1.0]{helvet}

\usepackage{newtxmath}

\usepackage{booktabs}
\usepackage{multirow}
\usepackage{makecell}
\renewcommand{\arraystretch}{1.0}

\usepackage{setspace}
\usepackage[font={normal,stretch=1.0},labelfont=bf]{caption}
\usepackage{subcaption}

\usepackage{siunitx}

\usepackage{amsmath}
\usepackage{amsfonts}
\usepackage[utf8]{inputenc}
\usepackage{url}
\usepackage{nicefrac}
\usepackage{microtype}
\usepackage{graphicx}
\usepackage{float}

\usepackage{hyperref}
\hypersetup{
  colorlinks=true,
  urlcolor={blue},
  citecolor={blue},
  linkcolor={black},
}
\usepackage[style=phys,
            biblabel=brackets,
            sorting=none,
            maxnames=3,
            minnames=1,
            doi=false,
            isbn=false,
            url=false,
            eprint=false]{biblatex}
\title{Context-Aware Deep Learning for Defect Classification in Atomic-Resolution STEM}

\usepackage{fancyhdr}
\renewcommand{\undertitle}{}

\hypersetup{
pdftitle={Context-Aware Deep Learning for Defect Classification in Atomic-Resolution STEM},
pdfsubject={Deep learning, Scanning transmission electron microscopy, Defects},
pdfauthor={Jiadong~Dan},
pdfkeywords={Deep learning, Scanning transmission electron microscopy, Defects},
}

\date{}

\begin{document}

\maketitle

\newcommand{\authorvspace}{-9em}

\vspace{\authorvspace}

\begin{center}
Jiadong Dan\textsuperscript{1,2,*},
Cheng Zhang\textsuperscript{3},
Leyi Loh\textsuperscript{4},
Ivan Verzhbitskiy\textsuperscript{5},
Yuan Chen\textsuperscript{6},
Goki Eda\textsuperscript{6,7},
Michel Bosman\textsuperscript{8},
N. Duane Loh\textsuperscript{1,2,6,$\dagger$}
\end{center}
\vspace{0.5em}

\vspace{-1em}
\begin{center}
\begin{minipage}{0.75\textwidth}
\small
\raggedright
\textsuperscript{1}Department of Biological Sciences, National University of Singapore, Singapore\\[0.3em]
\textsuperscript{2}Centre for BioImaging Sciences (CBIS), National University of Singapore, Singapore\\[0.3em]
\textsuperscript{3}Department of Materials Science and Engineering, City University of Hong Kong, China\\[0.3em]
\textsuperscript{4}Department of Materials Science and Metallurgy, University of Cambridge, United Kingdom\\[0.3em]
\textsuperscript{5}Quantum Innovation Centre (Q. InC), Agency for Science Technology and Research (A*STAR), Singapore\\[0.3em]
\textsuperscript{6}Department of Physics, National University of Singapore, Singapore\\[0.3em]
\textsuperscript{7}Department of Chemistry, National University of Singapore, Singapore\\[0.3em]
\textsuperscript{8}Department of Materials Sciences and Engineering, National University of Singapore, Singapore\\[0.3em]
\end{minipage}
\end{center}
\vspace{1em}

\renewcommand{\thefootnote}{}
\footnotetext{\textsuperscript{$*$, $\dagger$}Corresponding authors: jiadong.dan@u.nus.edu; duaneloh@nus.edu.sg}
\renewcommand{\thefootnote}{\arabic{footnote}}

\vfill

\newpage
\begin{abstract}
Artificial intelligence is rapidly advancing materials characterization, yet most applications in electron microscopy rely solely on image contrast, overlooking the chemical and experimental context that shapes those images. This limitation makes defect classification inherently ambiguous, as similar contrasts can arise from different materials or imaging conditions. Here we develop a context-aware learning framework that integrates image-derived contrast with metadata describing composition, beam energy, and detector geometry. Using a systematically constructed dataset of \textasciitilde{}55 million simulated patches spanning 576 simulated cases across 96 doped monolayer transition-metal dichalcogenides, we show that conditioning on contextual variables transforms defect classification from an ill-posed image-only task into a well-posed, physically grounded problem. The framework achieves over 98\% accuracy on simulations and near-human agreement on experimental data, with a 94\% reduction in posterior entropy. By emphasizing contextual grounding over architectural complexity, this approach links experimental image contrast to the underlying chemical and imaging conditions, supporting physically grounded defect assignments and a general pathway toward multimodal AI models for autonomous materials characterization.
\end{abstract}

\newpage

\section{Introduction}

In materials characterization, reliable interpretation of instrumental data depends not only on measured signals but also on contextual information about the sample's chemical composition and experimental conditions. In scanning transmission electron microscopy (STEM), various data modalities jointly reveal the structural and chemical details of materials at the atomic scale. High-angle annular dark-field (HAADF) imaging uses Z-contrast to visualize atomic structures~\cite{Krivanek2010-wc, Treacy2011-zx}, where heavier elements appear brighter, while annular bright-field (ABF) imaging is more effective for detecting lighter elements, such as hydrogen and oxygen~\cite{Findlay2010-gu}. Spectroscopic techniques, including electron energy-loss spectroscopy (EELS) and energy-dispersive X-ray spectroscopy (EDS), map the sample's chemical composition~\cite{Zhou2012-iu}, showing elemental distribution and providing insights into bonding environments~\cite{Keast2001-nw} and oxidation states~\cite{Tan2011-he}. Diffraction-based methods, such as four-dimensional STEM (4D-STEM)~\cite{Ophus2019-mn}, generate spatially resolved diffraction patterns to study strain~\cite{Shi2024-zh}, crystallographic phases~\cite{Ni2023-bk}, and symmetries~\cite{Hirata2011-an, Liu2013-cz} within the material. In addition to data from imaging and spectroscopy, details such as the sample's chemical composition and experimental settings, often stored as text, are crucial for understanding the context of the results. Together, these data types create a comprehensive toolkit for investigating material structures, compositions, and functional properties.

Despite the broad capabilities of these measurement modalities, most artificial intelligence (AI) applications in electron microscopy remain focused on single-modality tasks~\cite{Lee2022-tc, Lobato2024-jo, Wang2024-ns, Li2024-cc, Chen2023-rv}, such as defect classification using atomic resolution STEM images or manifold learning from 4D-STEM diffraction patterns~\cite{Li2019-hj}. This narrow focus contrasts with the approach of human microscopists, who synthesize information across multiple modalities when analyzing and interpreting data. However, the integration of contextual information remains limited in AI-driven microscopy, constraining both generalization and interpretability.

Neglecting contextual information may render inference problems fundamentally ill-posed, since distinct physical conditions can yield similar image observations; incorporating contextual variables restores a well-posed mapping (see Supplementary Section S1 for a formal definition). Formally, single-modality models attempt to learn an ambiguous mapping $f(x)\!\rightarrow\!y$, where $x$ represents the observed image contrast and $y$ the target label. By conditioning on auxiliary information $c$---such as chemical composition, beam energy, or detector configuration---the task becomes learning $f(x\mid c)\!\rightarrow\!y$, where contextual information grounds the learned relationships in the underlying physics. This phenomenon is not unique to microscopy: medical diagnosis from X-rays without patient history~\cite{Seah2021-sd}, weather prediction from satellite images without atmospheric profiles~\cite{Allen2025-lg}, and autonomous driving from cameras without LiDAR~\cite{Li2022-ew} all illustrate the same ambiguity. Conditioning on auxiliary information constrains the solution space, transforming an ill-posed mapping into a well-posed one.

Atomic defect classification in monolayer materials provides a natural test case for this principle, because identifying the structure, type, and distribution of defects is central to understanding local material properties, while the reduced dimensionality of monolayers makes individual atomic columns directly accessible to STEM. The contrast of atomic-resolution STEM images depends sensitively on both chemical composition and experimental conditions: identical defect types can appear different under varying beam energies or detector geometries, while distinct defects may exhibit similar intensities. For instance, simulated HAADF images of monolayer WTe\textsubscript{2} and NbSe\textsubscript{2} show nearly identical contrast at 80~keV (Fig.~S1), and the apparent brightness of Ta and Se\textsubscript{2} columns in monolayer TaSe\textsubscript{2} reverses between HAADF and medium-angle annular dark-field (MAADF) modes (Fig.~S2). These examples illustrate that image-based models trained in a fixed context capture correlations that are locally valid but fail to generalize across materials or different imaging setups.

To overcome these limitations, we establish a general framework for context-aware learning in electron microscopy, in which both image contrast and contextual metadata are jointly used during model training and inference. We demonstrate that explicitly conditioning on chemical and experimental variables transforms ill-posed mappings into well-posed ones, enabling models to generalize across diverse materials and imaging setups. Using a large, systematically constructed dataset of simulated STEM defects spanning 96 doped transition-metal dichalcogenides (MX\textsubscript{2}+D) under multiple beam energies and detector configurations, we show that conditioning models on experimental context substantially improves classification accuracy and reduces posterior entropy, meaning the uncertainty in the model's predicted class probabilities. This uncertainty reduction corresponds to increased information gain and enhances model interpretability and robust cross-material transferability. This work highlights contextual grounding as a unifying principle for data-driven discovery in microscopy and provides a pathway toward more generalizable and physically consistent AI models for materials characterization.



\begin{figure}
    \centering
    \includegraphics[width=0.90\textwidth]{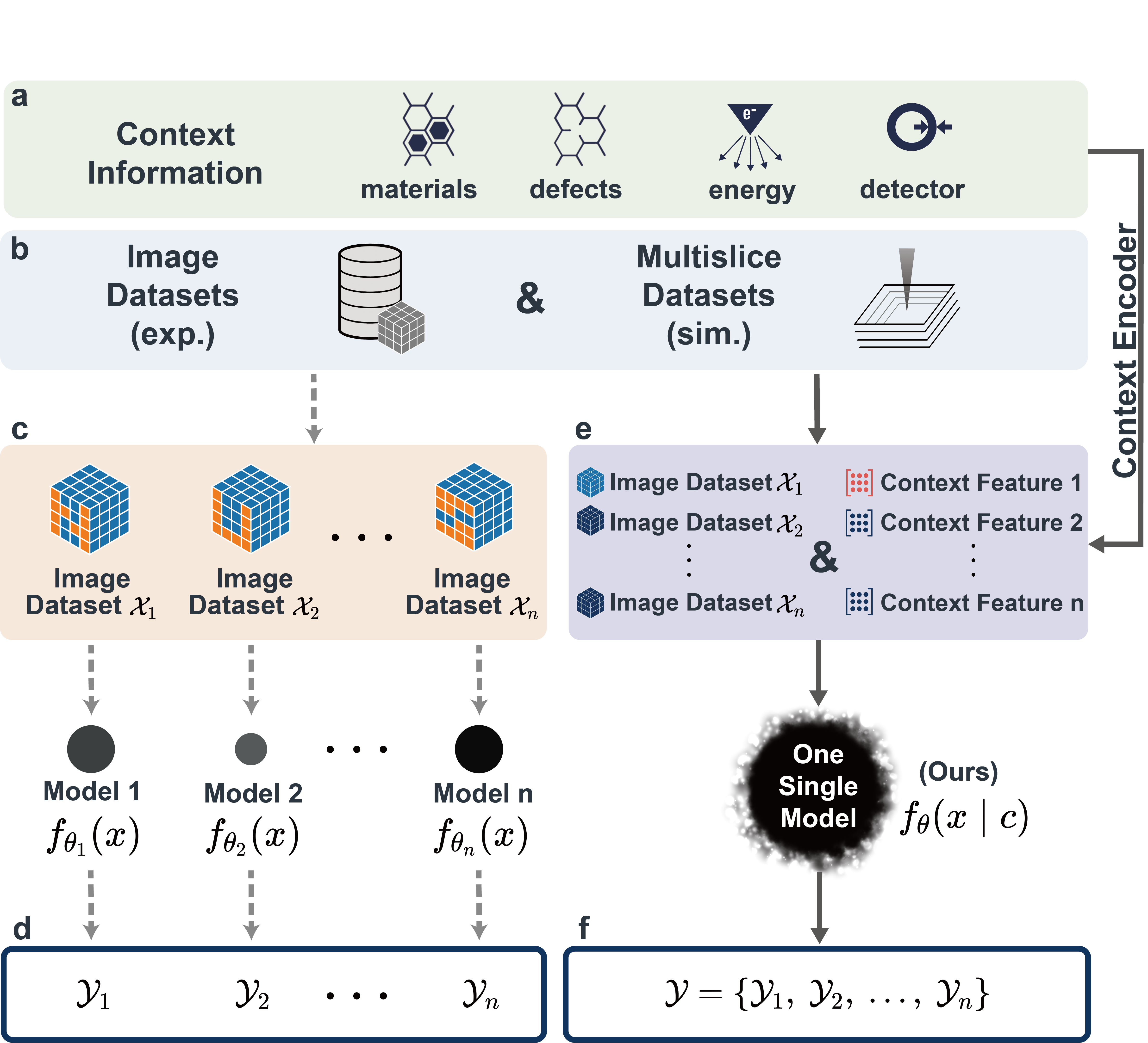}
    \caption{\textbf{From implicit to explicit context: a general framework for context-aware defect classification in electron microscopy.} \textbf{(b–d)} Conventional single-modality learning relies solely on atomic-resolution STEM images as input. Within such models, the mapping between image contrast and defect label is learned under a fixed and often implicit experimental context—defined by a particular material, beam energy, and detector configuration—limiting transferability across systems. \textbf{(a, e–f)} In our context-aware framework, chemical composition, beam energy, and detector geometry are explicitly included as contextual metadata. These variables are encoded as numerical context vectors and fused with image-derived features during training. Conditioning the classifier on contextual information transforms the ill-posed mapping $f(x)\!\rightarrow\!y$ into a well-posed mapping $f(x\mid c)\!\rightarrow\!y$, enabling a single, interpretable, and generalizable model for defect classification across diverse materials and imaging conditions.}
    \label{fig:fig_overview}
\end{figure}

\begin{figure}
    \centering
    \includegraphics[width=0.85\textwidth]{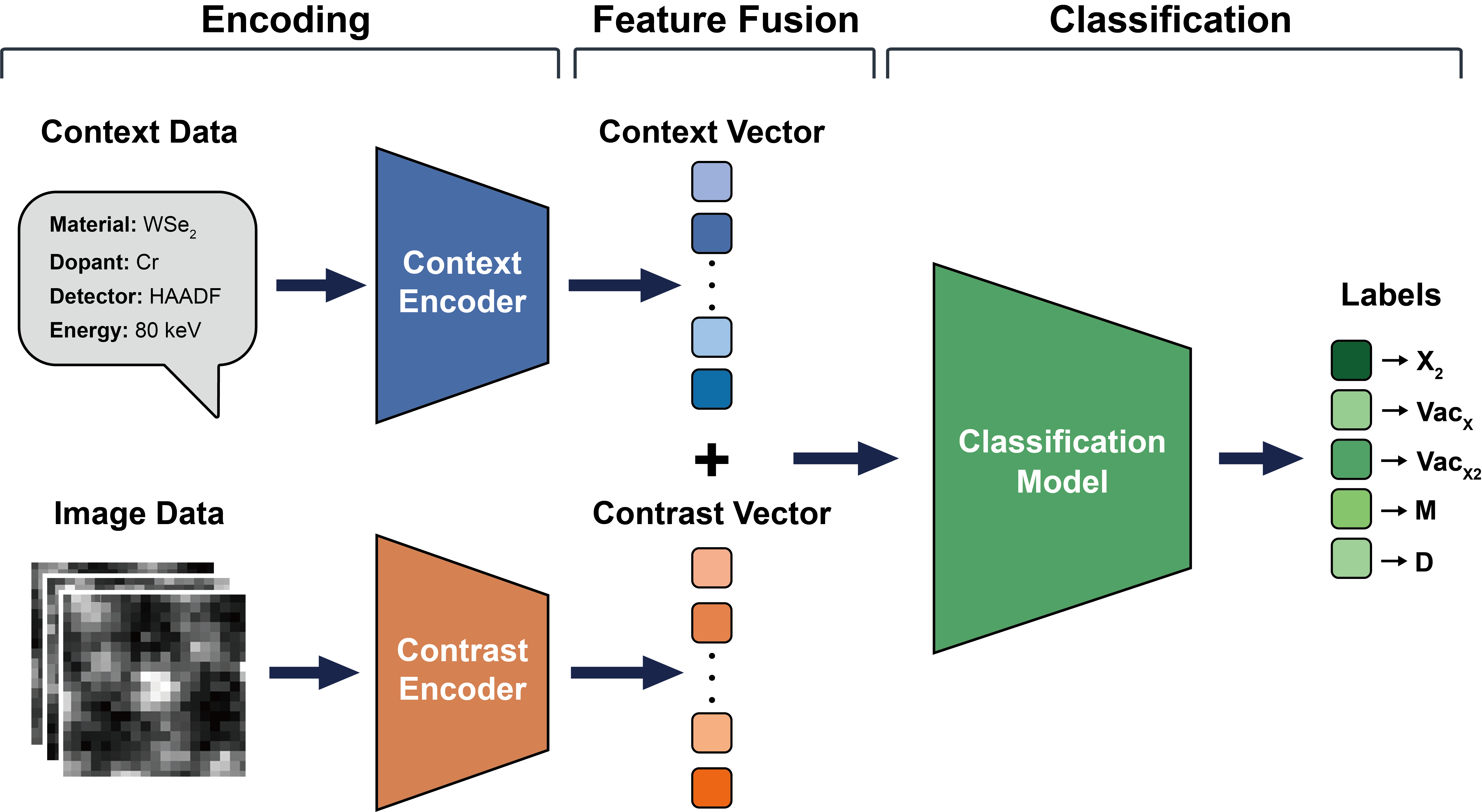}
    \caption{\textbf{The architecture of the overall learning framework comprising \textit{encoding, fusion, and classification}.} The context encoder (blue trapezoid) processes experimental context data, while the contrast encoder (orange trapezoid) processes image data, converting both into one-dimensional feature vectors. These vectors are concatenated and input into the classification model (green trapezoid), which categorizes them into one of five atomic column types: X\textsubscript{2}, single-vacancy, double-vacancy, metal, and dopant.}
    \label{fig:fig_multimodal}
\end{figure}

\section{A Framework for Context-Aware Defect Classification}

Figure~1 illustrates the conceptual foundation of context-aware learning for defect classification in electron microscopy. The left panels (Figs.~1b–d) show a conventional supervised approach in which the models only receive images as input. Within such image-only models, the mapping between image dataset $\mathcal{X}$ and defect label $\mathcal{Y}$ is learned under a fixed and often implicit experimental context---defined by a particular material, beam energy, and detector configuration. Such models achieve internal consistency but do not transfer reliably across different imaging conditions or materials. In contrast, the right panels (Figs.~1a, 1b, and 1e–f) illustrate the context-aware approach, in which both image data and contextual metadata are provided as inputs to the learned model $f_{\theta}(x\mid c)$ (with $\theta$ denoting the trainable parameters). By explicitly incorporating contextual variables into model training and inference, the mapping from image contrast to defect identity becomes physically grounded and generalizable across materials and imaging setups.

Figure~2 presents a general framework applicable to various classifier architectures. The framework consists of three key stages: \emph{encoding}, \emph{fusion}, and \emph{classification}. In the encoding stage, separate encoders process image and contextual inputs. The image encoder extracts spatial contrast features from atomic-resolution STEM images, while the context encoder transforms chemical and experimental descriptors into numerical embeddings. During feature fusion, the encoded vectors are combined to form a joint representation that captures characteristics of both contrast and context. The fused representation is then used by a classifier, such as a multilayer perceptron (MLP)~\cite{Rumelhart1986-xt} or transformer~\cite{Vaswani2017-mc}, to predict atom or defect types.

Encoders are essential because of the heterogeneous nature of the inputs: contextual information---including the sample’s chemical formula, beam energy, and detector geometry---differs fundamentally from spatially structured image data. Encoders transform this information into latent representations that can be combined meaningfully. Although end-to-end training of encoders and classifiers is common, it often yields latent vectors that are hard to interpret and require large datasets~\cite{Bengio2012-ny}. For scientific variables that are already well defined and physically interpretable, pre-designed features often suffice~\cite{Ward2018-mi}. In our work, we employ a predefined encoding scheme for chemical and experimental information. This approach enhances interpretability, reduces the number of learnable parameters, simplifies training, and lowers the demand for extensive datasets. As shown later, combining these encoders with multiple supervised architectures, including MLP and attention models, yields high accuracy, lower posterior entropy, and higher information gain once contextual information is properly integrated.

Having established the need for interpretable encoders, we now turn to how image and contextual information are transformed into compatible representations for fusion and joint learning.


\begin{figure}
    \centering
    \includegraphics[width=0.95\textwidth]{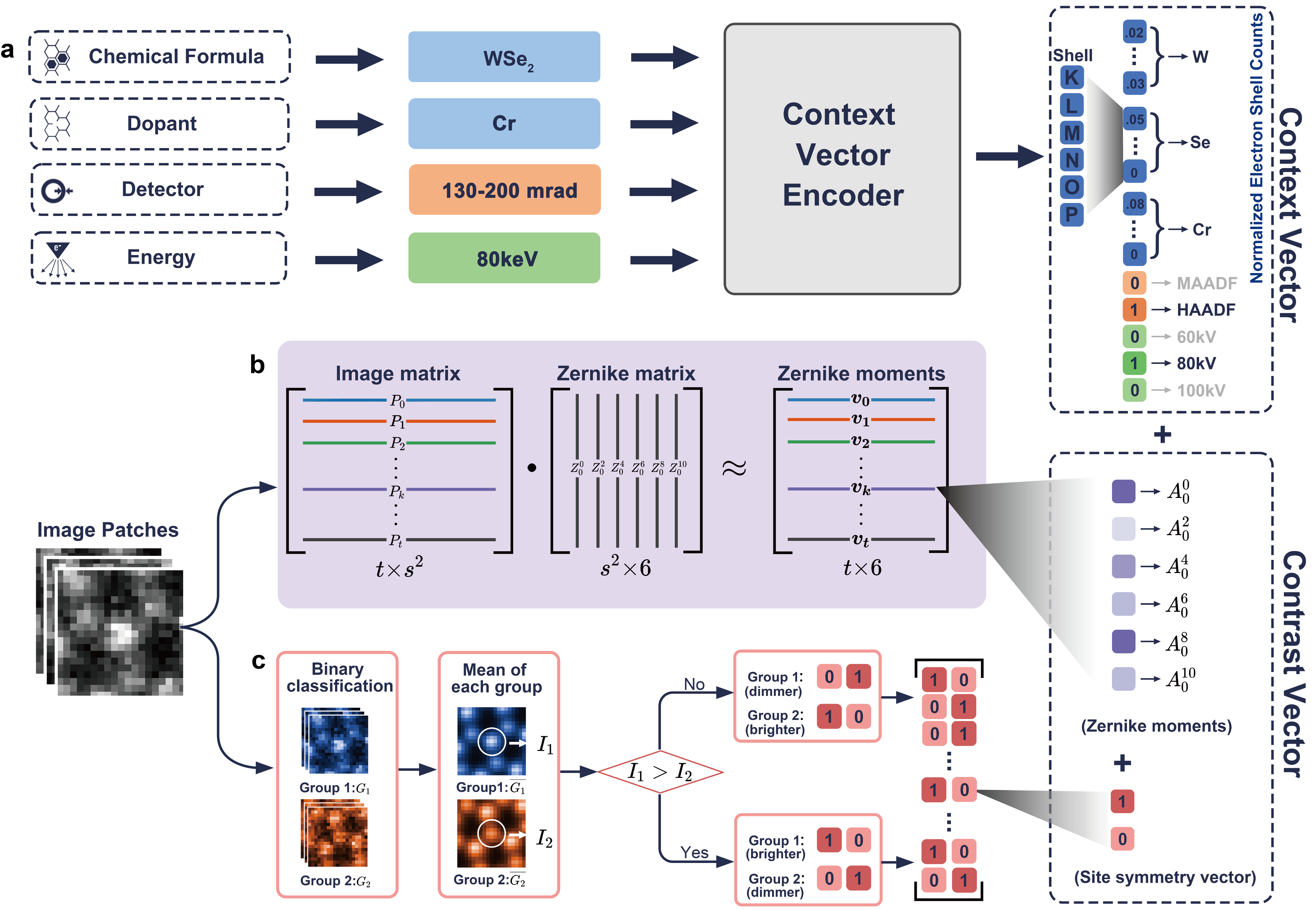}
    \caption{\textbf{Encoding of contextual and contrast information.} \textbf{(a)} Contextual metadata—including chemical composition, detector configuration, and electron-beam energy—is numerically encoded as feature vectors and concatenated sequentially to form the context vector. \textbf{(b)} Local STEM image patches are flattened and transformed into rotationally invariant Zernike moment features. \textbf{(c)} Patches are further categorized by site symmetry (metal or chalcogen sites) to compute site-symmetry vectors. The combined context and contrast features yield a fixed 31-dimensional representation for downstream model training.}
    \label{fig:fig_context}
\end{figure}

\section{Contextual and Local Image Patch Encoding}

To illustrate the encoding scheme, we select monolayer transition-metal dichalcogenides (MX\textsubscript{2}) with substitutional dopants (D) as a model material system (MX\textsubscript{2}+D; Fig.~S3). For this system, our key question is: What minimal contextual information is needed to distinguish materials when image contrast alone is insufficient? We propose that the metadata required by multislice simulations~\cite{Madsen2021-ok} to reproduce experimental STEM images defines this essential fingerprint. In the MX\textsubscript{2}+D systems, this fingerprint comprises four components: the MX\textsubscript{2} chemical formula, the dopant identity, the detector configuration, and the electron-beam energy. These descriptors are encoded into a single 23-dimensional context vector (Fig.~3a).

As a concrete example, we illustrate the encoding process using a WSe\textsubscript{2} system doped with chromium (Cr). In this case, the chemical fingerprint (WSe\textsubscript{2} + Cr) is represented by a vector of length 18 (blue), comprising three normalized electron shell configuration~\cite{Eickerling2008-uz} vectors of length 6 for W, Se, and Cr, respectively (Fig.~S4). Specifically, the configuration for W is denoted as {2, 8, 18, 32, 18, 8} and normalized by its atomic number, yielding {0.027, 0.108, 0.243, 0.432, 0.108, 0.027}. Similarly, the normalized vectors for Se and Cr are obtained. Detector types---HAADF and MAADF---are represented by a 2-element vector (orange), where each entry indicates the presence or absence of a specific detector. Electron beam energies (60 keV, 80 keV, 100 keV) are encoded as a 3-element vector (green), with each position corresponding to a specific beam energy. These individual vectors are concatenated into a context vector of length 23, containing all the essential information for downstream training.

In parallel, our contrast encoder operates on small, local STEM image patches rather than full-field images. Full-field multislice simulations are computationally impractical and would require petabyte-scale storage when variations in orientation, scale, noise, and defect density are fully sampled (Fig.~S5). While technically feasible, simulating and storing such a volume is not cost-effective. Restricting the input to local patches avoids this combinatorial explosion and keeps the training pipeline efficient and scalable.

Combining this patch-based strategy with Zernike encoding~\cite{Dan2022-to, Dan2024-ik} reduces data requirements by six orders of magnitude (Fig.~S5). The Zernike descriptors can be reformulated to be rotationally invariant, removing the need for orientation-based augmentation. Moreover, when patch size is chosen using a consistent scale-aware criterion, additional scale augmentation becomes unnecessary. Zernike features are also robust to diverse noise types~\cite{Dan2022-to, Dan2024-ik}, a property that promotes more reliable generalization under realistic imaging conditions.

Importantly, focusing on local patches also mitigates class imbalance. In wide field-of-view datasets, point defects often make up only a small fraction in the image~\cite{Loh2024-qf}, leading to biased learning~\cite{Johnson2019-ut, Chen2024-sj}. Our patch-based scheme avoids this problem by simulating patches centered on both defect sites and pristine matrix atoms, ensuring a well-balanced representation of all atomic species. This design keeps our pipeline both efficient and reliable.

Figure~3b shows how we extract Zernike features from each local patch centered on one atomic column and its three nearest neighbors. We compute rotationally invariant Zernike moments and then keep only those with azimuthal index m = 0 and radial index n $\leq$ 10, yielding a 6-dimensional feature vector (purple). 

Although Zernike features in principle encode enough contrast information to distinguish intensity differences between atomic columns, noise can blur the contrast boundaries that differentiate atom species. To sharpen feature separation in our MX\textsubscript{2} + D system, we add a two-dimensional site-symmetry descriptor that identifies the nominal sublattice of each patch: one component flags metal-centered sites, which have D\textsubscript{3}h local symmetry in ideal 1H MX\textsubscript{2}, and the other flags chalcogen-centered sites, which have C\textsubscript{3}v local symmetry. To implement this, we partition patches into two categories corresponding to the crystal’s two unique site symmetries (Fig.~3c). Averaging each group’s central‐atom intensities reveals a consistent intensity difference, which we encode as [1, 0] for the brighter group and [0, 1] for the dimmer group. Zernike and site symmetry features together form the contrast vector. 

Finally, we concatenate the context vector and the contrast vector to form the complete features for downstream model training and inference.


\begin{figure}
    \centering
    \includegraphics[width=0.95\textwidth]{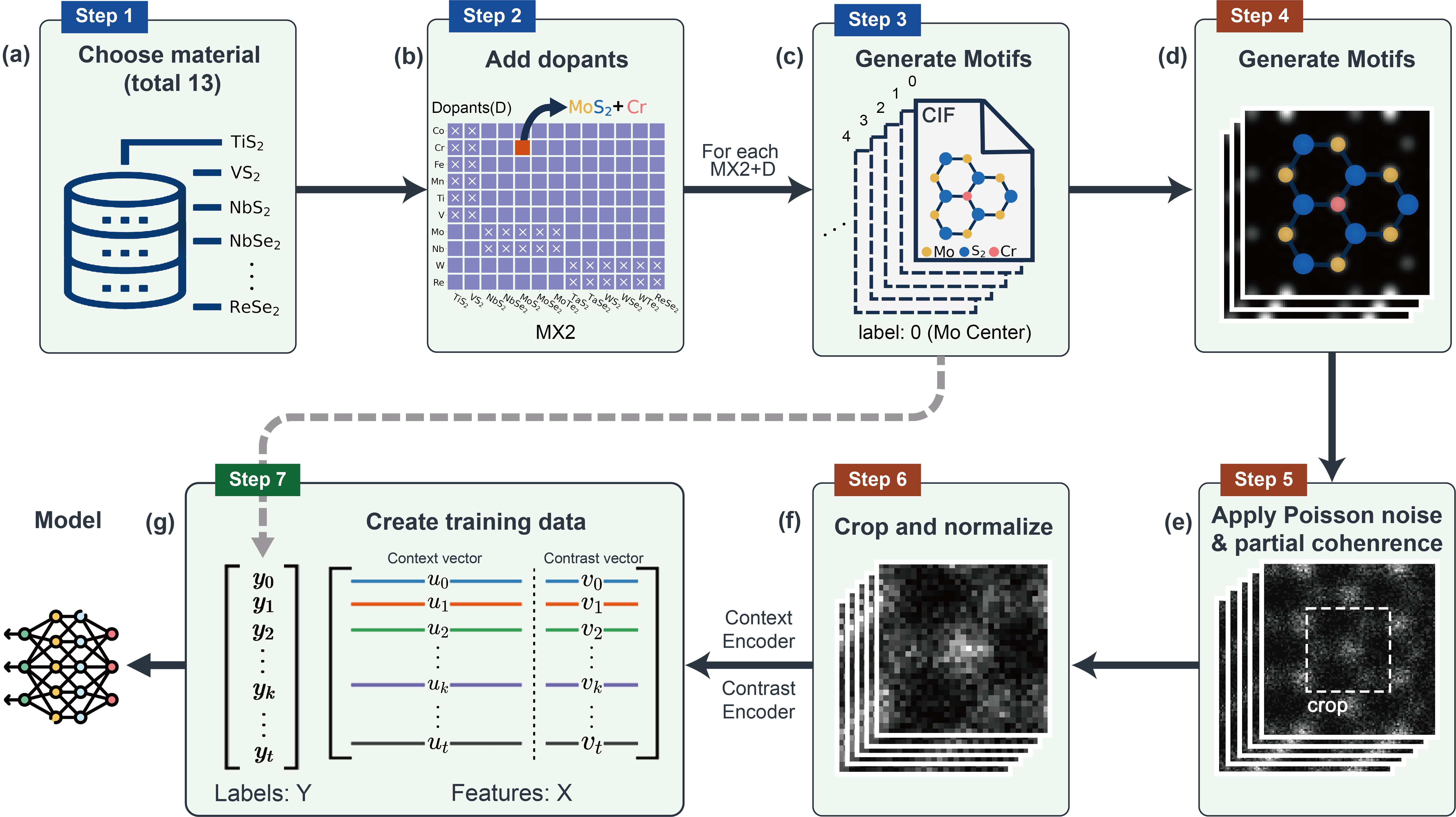}
    \caption{\textbf{Workflow for Generating Simulation Training Datasets}. \textbf{(a)} We select 13 existing 1H transition metal dichalcogenides (TMD) materials (MX2). We ensure their existence and record lattice parameters from the Materials Project. \textbf{(b)} We identify 10 different types of dopants from published sources. By combining these dopants with the materials, we create 96 distinct MX2+dopant combinations.  \textbf{(c)} For each MX2+dopant combination, we generate five different structural motifs (metal, dopant, X2, single vacancy, double vacancy) with labels, stored as Crystallographic Information Files (.cif); the full set of local motif configurations is shown in Fig.~S7. \textbf{(d)} Using a multislice simulation algorithm, we generate simulated images for each motif. \textbf{(e)} To reduce out-of-distribution between simulated and experimental images, we add various types of noise, including Poisson, Gaussian, and scan noise. \textbf{(f)}  We crop all simulated images according to a standard that ensures each image includes the central atom columns and three surrounding neighboring atoms. These images are then transformed into Zernike Moments. \textbf{(g)} The images are encoded Zernike Moments and concatenated with context information into context vectors, forming the input features X. The labels from panel c serve as the output labels Y. Together, these form the training data for model training.}
    \label{fig:fig_dataset}
\end{figure}

\section{Dataset Preparation}

We curated an extensive dataset of 1H-phase monolayer transition-metal dichalcogenides (MX\textsubscript{2}) doped with a variety of transition-metal species (D). This dataset serves as a consistent testbed for evaluating how contextual variables influence model generalization across materials and imaging conditions. The stepwise data-curation pipeline is outlined in Fig.~4.

First, we conducted an exhaustive search of the Materials Project database~\cite{Horton2025-of} and identified 13 experimentally realized 1H monolayer transition metal dichalcogenides (Fig.~4a). From these hosts, we selected 10 candidate dopants, resulting in 96 unique MX\textsubscript{2}+D combinations (Fig.~4b and Fig.~S6). To ensure each dopant remains visible in atomic resolution STEM images, we excluded cases where the host metal and dopant occupy the same row of the periodic table, as their atomic numbers are too close to provide sufficient contrast in Z-contrast imaging. For example, Nb (Z = 41) in MoS\textsubscript{2} (Z = 42 for Mo) would be nearly indistinguishable in a dark-field STEM image especially under noisy conditions~\cite{Lebeau2008-vs}.

For each MX\textsubscript{2}+D combination, we define five local atomic arrangements according to the central column identity: dichalcogen center (X\textsubscript{2}), single X vacancy (Vac\textsubscript{X}), double X vacancy (Vac\textsubscript{X2}), metal center (M), and dopant center (D), assigned class labels 0 through 4 respectively. These local structures are labeled from 0 to 4. Within each class, we hold the central column fixed and systematically vary its neighbors, yielding 34 unique local environments (Fig.~S7). We then generate a Crystallographic Information File (CIF) for each arrangement to drive our multislice simulations~\cite{Madsen2021-ok}. We use idealized local motifs with fixed lattice positions to provide a controlled benchmark in which the effects of chemical identity and imaging context on STEM contrast can be isolated systematically; robustness to real structural deviations, including local relaxations, is assessed through experimental validation. This local approach ensures a balanced dataset, avoiding the class imbalance typically found in full-field STEM images where dopants and vacancies are rare~\cite{Loh2024-qf} (Fig.~4c).

We then simulated STEM images using the multislice algorithm for all 34 local configurations (Fig.~4d). For each MX\textsubscript{2}+D system, simulations are conducted at three electron-beam energies and two detector geometries to capture a range of imaging conditions, resulting in a total of 576 distinct contextual cases. To emulate experimental variations, each image is further augmented with probe-size–dependent Gaussian blur, Poisson noise, and scan distortion (Fig.~4e). The corresponding noise and augmentation parameters are summarized in Fig.~S8.

To prepare image patches for the contrast encoder, we cropped each simulated image according to a fixed workflow (Fig.~S9), ensuring that it includes the central atomic column and its three nearest neighbors, as illustrated in Fig.~4f.

Following the encoding pipeline in Fig.~3, each patch is converted into a contrast vector and its metadata encoded into a context vector. We concatenate these vectors into our feature set $\mathbf{X}$, while the five structure labels from Fig.~4c form the targets $\mathbf{Y} \in \{0,1,2,3,4\}$. The resulting $(\mathbf{X}, \mathbf{Y})$ pairs make up the training and testing dataset (Fig.~4g).

The final dataset occupies approximately 6.8~gigabytes (Fig.~S5), providing a compact yet information-rich foundation for training our unified defect classification model. In contrast, training an image-to-image translation model across all MX\textsubscript{2}+D configurations would require a dataset on the order of 6~petabytes (Fig.~S5), underscoring the necessity and efficiency of our patch-based approach. Although our simulated images incorporate multiple types of noise to mimic experimental conditions, a domain gap still exists. The Zernike-based contrast representation helps mitigate this gap, as Zernike moments are inherently robust to noise and imaging aberrations~\cite{Dan2022-to}.


\begin{figure}
    \centering
    \includegraphics[width=0.95\textwidth]{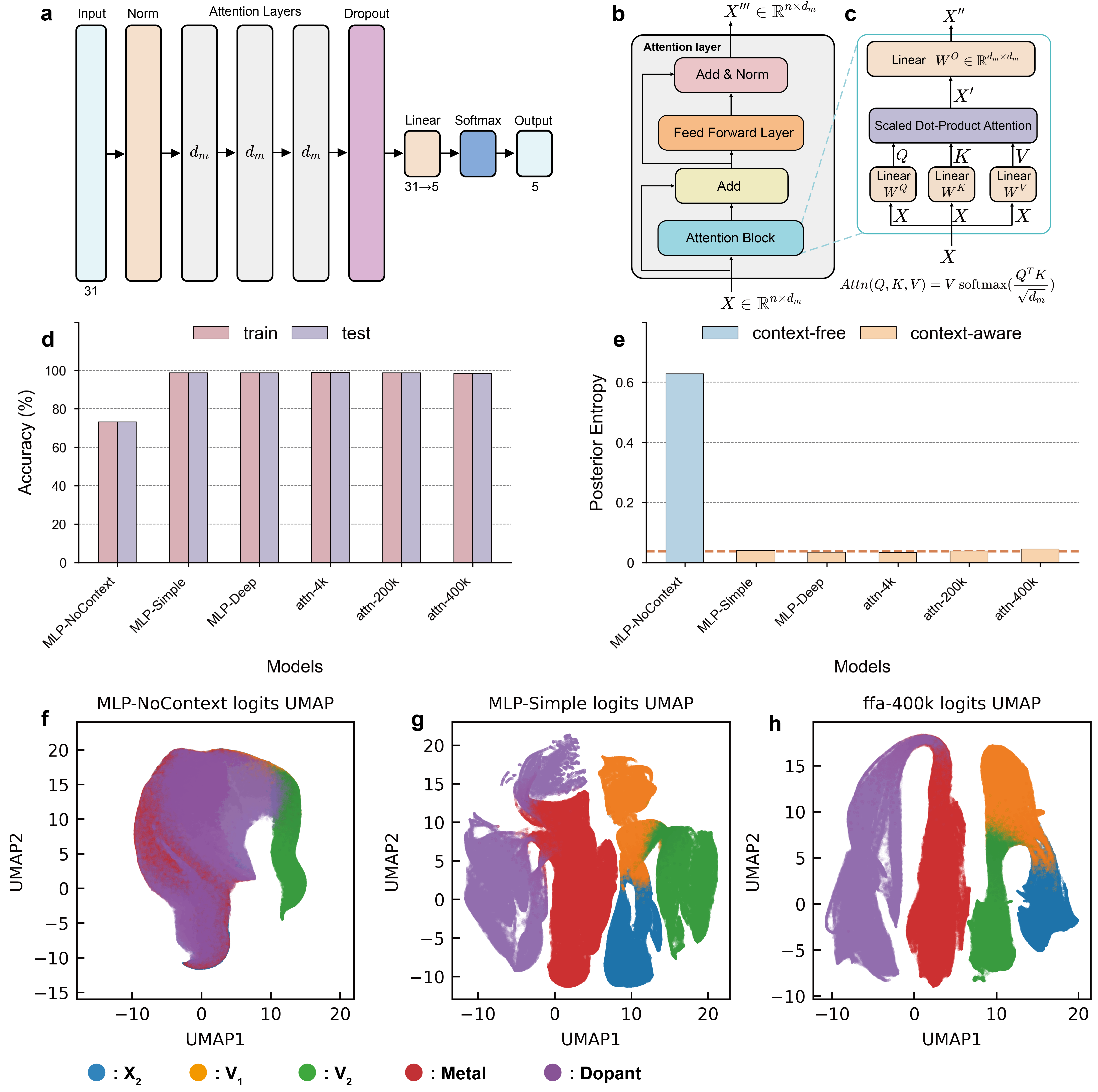}
    \caption{\textbf{Architecture and performance of defect attention model}. 
    \textbf{(a)} Schematic of the attention model with three stacked attention layers and classification head. 
    \textbf{(b)} Attention layer components: self-attention (cyan) and feed-forward (orange) sub-layers with residual connections and batch normalization. 
    \textbf{(c)} Attention block detail showing linear projections to queries (Q), keys (K), and values (V), followed by scaled dot-product attention. 
    \textbf{(d)} Train and test accuracies for six models: context-free MLP-NoContext, context-aware MLPs (MLP-Simple, MLP-Deep), and attention models (attn-4k, attn-200k, attn-400k). 
    \textbf{(e)} Posterior entropy comparison showing approximately 94\% information gain with contextual inputs. 
    \textbf{(f-h)} UMAP visualizations: (f) overlapping input features without context; (g,h) well-separated learned representations with context (MLP-Simple, attn-400k).
    }
    \label{fig:fig_attention_model}
\end{figure}

\section{Model Training}

Based on our curated MX\textsubscript{2}+D dataset, we trained six classifiers to identify point defects. Three attention-based models (attn-4k, attn-200k, and attn-400k) were trained with batch sizes of 4,096, 204,800, and 409,600, respectively. Three MLP baselines with varying architectural complexity were also trained: MLP-NoContext (4 hidden layers, trained without contextual information, Fig.~S10), MLP-Simple (4 hidden layers, Fig.~S11), and MLP-Deep (11 hidden layers, Fig.~S12). Except for MLP-NoContext, all models were trained on the same dataset with identical contextual features and evaluated on the same five-class classification task.

The attention-based architecture~\cite{Vaswani2017-mc} (Fig.~5a) takes a 31-dimensional input consisting of encoded context and contrast features. The input is first processed with batch normalization. The model comprises three identical attention layers (Fig.~5b), each preserving the feature dimension and consisting of a feature--feature self-attention block (Fig.~5c) followed by a position-wise feed-forward layer. Unlike the original Transformer, which employs layer normalization~\cite{Ba2016-rr}, we adopt batch normalization~\cite{Ioffe2015-hz}, as our inputs consist of independent feature vectors rather than correlated token sequences. A dropout rate of 0.1 is applied after the final attention block. Finally, a classification head, composed of a linear layer and softmax activation, maps the refined features to five output classes. The three attention variants differ only in batch size (4\,096; 204\,800; and 409\,600).

To leverage correlations across feature dimensions rather than sample sequences, we implemented feature--feature attention instead of the conventional token--token mechanism. A detailed view of the feature--feature attention block is shown in Fig.~5c. The input features are given by $X \in \mathbb{R}^{n \times d_m}$, where $n$ is the number of samples and $d_m$ is the feature dimension. We project $X$ into queries, keys, and values $Q$, $K$, $V \in \mathbb{R}^{n \times d_m}$.
Scaled dot-product attention is then computed as
\[
\mathrm{Attention}(Q,K,V) 
= V \, \mathrm{softmax}\!\left(\frac{Q^{T}K}{\sqrt{d_m}}\right),
\]
where $Q^{T}K \in \mathbb{R}^{d_m \times d_m}$ encodes pairwise correlations between column features. The resulting output vectors $X^{\prime\prime} \in \mathbb{R}^{n \times d_m}$ are passed through a final linear layer to produce the attention output $X^{\prime\prime\prime} \in \mathbb{R}^{n \times d_m}$.

For comparison, we implemented two MLP baselines: MLP-Simple and MLP-Deep~\cite{Hornik1989-yg, Rumelhart1988-uy}. MLP-Simple consists of three sequential blocks of linear–ReLU–dropout layers~\cite{Agarap2018-ff, Srivastava2014-yz}, with hidden dimensions 64, 32, 16, and 8 across the layers. A final linear layer followed by a softmax activation maps the features to output classes. MLP-Deep is a deeper variant with 12 such blocks before the classification head.

Train and test accuracies of all models are summarized in Fig.~5d. MLP-NoContext plateaued at both training and test accuracies of approximately 73\% , indicating substantial overlap among features. This overlap is further illustrated by the Uniform Manifold Approximation and Projection (UMAP)~\cite{Sainburg2021-ap} visualizations of the input features and pre-softmax logits (Fig.~5f and Fig.~S13).

Incorporating contextual information yields consistently high performance across all architectures, with test accuracies ranging from 98.36\% to 98.71\%. The attention-based models (attn-4k, attn-200k, attn-400k) achieve 98.71\%, 98.53\%, and 98.36\%, respectively, while MLP-Simple and MLP-Deep both attain 98.59\%. UMAP visualization of the learned feature spaces (Fig.~5g,h) shows that context-aware models, exemplified by MLP-Simple and attn-400k, produce well-separated class clusters with minimal ambiguity, in contrast to the no-context baseline (additional UMAP layouts in Fig.~S14).

We further quantified the information gain provided by contextual information using posterior entropy~\cite{MacKay1992-mh, Hennig2011-zp} (Fig.~5e). The context-free model (MLP-NoContext) yields an entropy of 0.628 nats, whereas context-aware models exhibit an average entropy of 0.0375 nats (orange dashed line). This corresponds to an information gain of approximately 94\%, indicating a pronounced reduction in posterior uncertainty. For a five-class problem, the theoretical maximum entropy is 1.61 nats; context-aware models operate at only 2.3\% of this maximum, demonstrating that contextual cues effectively resolve classification ambiguity.


Calibration was assessed using the Expected Calibration Error (ECE)~\cite{Pakdaman-Naeini2015-rk, Guo2017-lo}. Both context-free and context-aware models exhibit low ECE values (Fig.~S15), indicating reliable confidence estimates. However, incorporating contextual features not only improves accuracy from 73\% to over 98\% but also shifts the confidence distribution toward high certainty (peak near 1.0, Fig.~S16). Thus, contextual information resolves class ambiguities while preserving calibration, making the model both more accurate and more confident in its correct predictions.

Although the MLP baselines achieved comparable accuracies when provided with contextual inputs, the attention architecture offers an additional layer of interpretability by revealing which contextual and contrast features the model emphasizes during classification. To elucidate how contextual information enhances performance, we examined the attention maps of the attention models (Fig.~S17). The third-layer attention maps display pronounced vertical stripes, indicating that specific Key features are attended to by nearly all Query features. These globally influential features function as a learned, high-level context that modulates how local contrast-based features are interpreted. This pattern suggests that the network has acquired a hierarchical representation in which contextual information acts as a shared reference for feature interactions, aligning with the physical intuition that defect contrast depends on imaging conditions.


\begin{figure}
    \centering
    \includegraphics[width=0.95\textwidth]{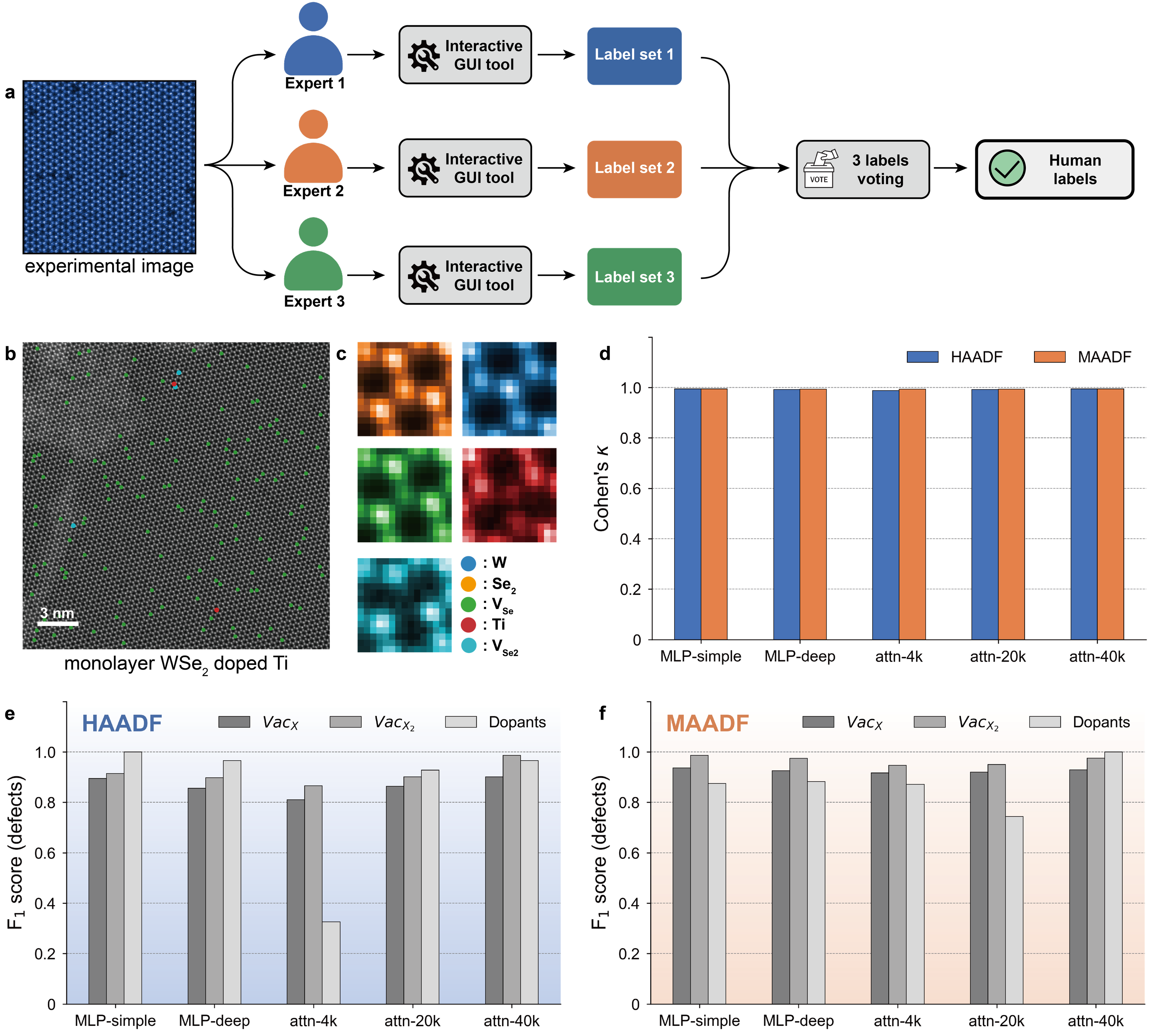}
    \caption{\textbf{Workflow and evaluation of human and model classifications on a multi-dopant WSe\textsubscript{2} HAADF/MAADF image dataset}. \textbf{(a)} Three experts independently labeled each image via an interactive GUI; majority voting produced the reference consensus label. \textbf{(b)} Representative HAADF image of Ti-doped WSe\textsubscript{2} showing human-identified defects: single Se vacancies (green), double Se vacancies (cyan), and Ti dopant sites (red). \textbf{(c)} Mean image patches for the five atomic classes—W sites, Se sites, single Se vacancy sites, double Se vacancy sites, and Ti dopant sites—averaged over all occurrences in the example image from panel (b). \textbf{(d)} Cohen’s $\kappa$ scores, averaged across all images, quantify agreement between human consensus labels and predictions from five model variants (MLP-Simple, MLP-Deep, attn-4k, attn-200k, and attn-400k) on both HAADF and MAADF. Scores near 1 indicate high consistency. \textbf{(e)} HAADF F\textsubscript{1} scores for the three defect classes across all models, highlighting attn-400k as the top performer. \textbf{(f)} MAADF F\textsubscript{1} scores for the same categories, showing a matching ranking.}
    \label{fig:fig_model_performance}
\end{figure}

\section{Model Performance on Experimental Data}

To evaluate our models on experimental data, we collected HAADF and MAADF STEM images of WSe\textsubscript{2} doped with Ti, V, Mn and Co respectively, and had three experts independently label each image patch. Each atomic site in the images was assigned one of five possible class labels: W atom, Se\textsubscript{2} atoms, Se single-vacancy, Se double-vacancy, or dopant. Final reference labels were determined via majority voting across annotators (Fig.~6a).

With these consensus labels in hand, we then projected the annotated defects back onto the original micrographs to visualize their spatial distribution. Figure~6b highlights the mapped locations of single and double Se vacancies alongside Ti dopants in monolayer WSe\textsubscript{2}. Figure~6c shows, for the same image, the mean patch of each structural class (W site, Se\textsubscript{2} site, Se single-vacancy, Se double-vacancy, Ti dopant), averaged over all occurrences.

We then measured agreement between human annotators and our models using Cohen’s $\kappa$, a statistic that corrects for agreement expected by chance~\cite{Cohen1960-rp, McHugh2012-oy}. Since our five labels {0, 1, 2, 3, 4} are nominal categories with no intrinsic ordering, we used the unweighted version of Cohen’s $\kappa$. Scores were computed for each of the five context-aware model variants—three attention-based models (attn-4k, attn-200k, attn-400k) and two MLP baselines (MLP-Simple and MLP-Deep)—and averaged across all test images for both HAADF and MAADF inputs. Figure~6d shows the Cohen's $\kappa$ for each model. Models with $\kappa$ close to 1 indicate strong alignment with the expert consensus~\cite{Landis1977-yi}.

While Cohen's $\kappa$ offers a general measure of accuracy, it treats all label types equally~\cite{Byrt1993-do}. Given that atomic defects are both rare and critical to materials' properties~\cite{Loh2024-qf}, we performed a focused evaluation of model performance in detecting defects specifically. To this end, we recast the five-class annotations into a binary scheme~\cite{He2009-rj}: defect (classes 1, 2, 4: Se single-vacancy, Se double-vacancy, Ti dopant) and non-defect (classes 0, 3: W and Se\textsubscript{2} atoms). This ensures our evaluation metrics reflect sensitivity to the minority class rather than being dominated by the majority of structurally regular sites.

For each model, we computed precision and recall on the recast binary labels~\cite{Powers2020-pl}. Precision represents the conditional probability that a site predicted as defective by the model was also labeled as defective by the human experts, while recall quantifies the conditional probability that a defect labeled by humans was correctly predicted by the model. These two metrics were then combined into the F\textsubscript{1} score~\cite{Christen2024-mn}, the harmonic mean of precision and recall, providing a balanced measure of performance. We report F\textsubscript{1} scores separately for HAADF and MAADF inputs and across all three defect classes (Fig.~6e and Fig.~6f).

As shown in Table 1, the F\textsubscript{1} scores show strong and consistent performance for single and double vacancies across all models, with values typically above 0.85 for both HAADF and MAADF inputs. The dopant class, however, displays greater variability: while MLP-Simple, MLP-Deep, and attn-400k maintain F\textsubscript{1} scores close to or equal to 1.0, attn-4k exhibits a pronounced decrease under HAADF imaging. Among the attention models, attn-400k provides the most balanced performance across defect types and imaging modes, whereas the smaller-batch variants (attn-4k and attn-200k) show less stable dopant recognition. These trends indicate that larger batch sizes encourage the network to learn representations that transfer more reliably across HAADF and MAADF contrasts, improving consistency in identifying chemically distinct defect configurations.

All models experience performance degradation when transferred from simulated to experimental images, but the magnitude varies systematically with batch size. We observe that larger-batch models exhibit smaller performance drops, suggesting that the more stable gradient estimates from larger batches may encourage learning of features that are less specific to simulation artifacts and more transferable to experimental data. Consequently, attn-400k shows the smallest performance decline, demonstrating superior resilience to the simulation–experiment domain shift.

By combining Cohen’s $\kappa$ for overall agreement with a class-collapsed F\textsubscript{1} focused on defects, our evaluation shows that our models not only align with expert consensus labels but also excel at identifying rare, physically meaningful defect sites (also see Fig.~S18).

\begin{table}[ht]
  \centering
  \scriptsize
  \setlength{\tabcolsep}{3pt} 
  \renewcommand{\arraystretch}{1.1} 

  \begin{tabular}{@{}l cc cc cc cc cc@{}}
    \toprule
    \multirow{2}{*}{Models}
      & \multirow{2}{*}{\makecell{Train Acc.\\(\%)}}
      & \multirow{2}{*}{\makecell{Test Acc.\\(\%)}}
      & \multicolumn{2}{c}{Cohen’s $\kappa$}
      & \multicolumn{2}{c}{$F_{1}$ (single vac.)}
      & \multicolumn{2}{c}{$F_{1}$ (double vac.)}
      & \multicolumn{2}{c@{}}{$F_{1}$ (dopant)} \\[-2pt]
      &  & 
      & {\scriptsize HAADF} & {\scriptsize MAADF}
      & {\scriptsize HAADF} & {\scriptsize MAADF}
      & {\scriptsize HAADF} & {\scriptsize MAADF}
      & {\scriptsize HAADF} & {\scriptsize MAADF} \\
    \cmidrule(lr){4-5} \cmidrule(lr){6-7} \cmidrule(lr){8-9} \cmidrule(lr){10-11}
    MLP-NoContext & 73.2 & 73.2 & / & / & / & / & / & / & / & / \\
    MLP-Simple    & 98.6 & 98.6 & .995 & .995 & .894 & .937 & .914 & .987 & 1.00 & .875 \\
    MLP-Deep      & 98.6 & 98.6 & .992 & .994 & .856 & .926 & .899 & .974 & .966 & .882 \\
    attn-4k       & 98.7 & 98.7 & .988 & .993 & .810 & .917 & .866 & .947 & .326 & .872 \\
    attn-200k     & 98.5 & 98.5 & .992 & .993 & .864 & .921 & .901 & .950 & .929 & .744 \\
    attn-400k     & 98.4 & 98.4 & .995 & .994 & .901 & .929 & .987 & .976 & .966 & 1.00 \\
    \bottomrule
  \end{tabular}

  \caption{Model performance summary showing training/testing accuracy (in \%), Cohen’s $\kappa$, and $F_{1}$ scores for single and double vacancies and dopants under HAADF and MAADF imaging conditions.}
\end{table}

\section{Discussion and Conclusion}

The integration of contextual information—chemical composition, beam energy, and detector geometry—fundamentally reshapes defect classification in atomic-resolution STEM. When models receive only image contrast, they face an ill-posed learning problem: distinct physical scenarios can produce similar image signatures, creating intrinsic ambiguity in the mapping $f(x) \rightarrow y$. By explicitly conditioning on context, this transforms into $f(x, c) \rightarrow y$, where relevant physical conditions guide the model toward unique solutions. This shift is reflected quantitatively in our results: posterior entropy drops from 0.628 nats (context-free) to 0.0375 nats (context-aware), representing a 94\% reduction in classification uncertainty. Context-aware models operate at only 2.3\% of the theoretical maximum entropy for five-class classification, demonstrating that contextual cues effectively collapse the hypothesis space and eliminate ambiguity. This uncertainty reduction translates directly into performance gains, with classification accuracy improving from 73\% to over 98\%, while models maintain excellent calibration and achieve near-human agreement on experimental data (Cohen's $\kappa \approx 0.99$). Importantly, these improvements arise not from architectural sophistication but from how information is encoded and presented to the model. Both simple MLPs and attention-based architectures converge to similar performance once context is provided, underscoring that robust generalization emerges primarily from proper problem formulation rather than model complexity.

While MLPs and attention models achieve comparable accuracy with contextual inputs, the attention architecture offers additional scientific value through interpretability. Feature--feature attention maps reveal which experimental variables drive specific predictions, providing insight into the physical principles learned by the model. Third-layer attention maps display pronounced vertical stripes, indicating that specific key features—representing contextual information—are attended to globally across all query features. This pattern suggests the network has learned a hierarchical representation where context acts as a shared reference frame that modulates how local contrast features are interpreted, aligning naturally with the physical intuition that defect contrast depends on imaging conditions. This demonstrates that the model has internalized domain-relevant structure rather than merely memorizing statistical correlations.

Our experimental validation on multi-dopant WSe\textsubscript{2} reveals systematic patterns in how models transfer from simulation to real data. All models experience performance degradation during this domain shift, but the magnitude varies with training batch size. Empirically, we observe that larger-batch models (attn-400k) exhibit smaller performance drops compared to smaller-batch variants (attn-4k), particularly for dopant classification where F1 scores vary from 0.326 to 1.00 depending on batch size and imaging mode. This suggests that training dynamics influence the types of features learned: larger batches may promote more stable convergence to generalizable representations that are less tied to simulation-specific artifacts. However, the mechanisms underlying this relationship warrant further investigation, as the interaction between batch size, domain adaptation, and feature learning remains an open question in deep learning theory.

This work addresses a fundamental gap between how expert microscopists reason—jointly considering structure, chemistry, and experimental setup—and how machine learning models typically process data. By formalizing multimodal reasoning through conditional learning, we provide a scalable foundation for AI models that generalize across materials and imaging conditions. The framework naturally extends beyond point defect classification to interface analysis, phase segmentation, strain mapping, and integration of multimodal signals such as EELS, EDS, and 4D-STEM diffraction patterns. More broadly, context-aware learning offers a pathway toward scientific foundation models that bridge understanding with experimental control. In autonomous microscopy workflows, models must not only interpret images but also reason about which experimental parameters to adjust based on sample properties and measurement goals. By treating experimental metadata as first-class inputs alongside images, our framework enables models to learn the coupled relationships between sample characteristics, instrumental settings, and observable contrast—knowledge that is essential for closed-loop optimization and autonomous characterization.

While our framework demonstrates strong performance on simulated data and promising transfer to experimental images, several challenges remain. The simulation-to-experiment domain gap, though reduced through noise augmentation and Zernike encoding, still causes measurable performance drops. Future work should explore domain adaptation techniques, physics-informed regularization, and active learning strategies that leverage small amounts of experimental data to bridge this gap more effectively. Additionally, extending the framework to handle continuous experimental parameters (e.g., defocus, thickness) rather than discrete categories would increase flexibility and enable more fine-grained control.

By explicitly incorporating experimental context into defect classification, we transform an ill-posed imaging problem into a well-posed multimodal learning task. This approach achieves over 98\% accuracy on simulated data, near-human agreement on experimental images, and 94\% reduction in posterior entropy—demonstrating that contextual grounding is not merely beneficial but essential for robust AI in materials characterization. Our framework establishes a general paradigm for physics-consistent machine learning in electron microscopy, providing a foundation for autonomous characterization systems that can reason about the coupled relationships between sample properties, instrumental parameters, and observable contrast. As AI becomes increasingly central to materials discovery, context-aware learning offers a principled path toward models that are interpretable, generalizable, and aligned with the multimodal reasoning of expert scientists.

\section{Method}

\subsection{Conditional learning}

We denote atomic-resolution STEM images by $x \in \mathcal{X}$ and defect labels by $y \in \mathcal{Y}$. Contextual metadata, including chemical composition, dopant type, beam energy, and detector geometry, are represented collectively as $c$. A neural-network model with trainable parameters $\theta$ is written as $f_{\theta}(x \mid c)$. When required, the model outputs are interpreted as class probabilities via the softmax of the network logits.

\subsection{TMD experiments}

WSe\textsubscript{2} bulk crystals with various dopants were grown with the self-flux method~\cite{Vu2023-fw}. Single crystals were harvested from the bulk crystals and thinned down to monolayer thickness through mechanical exfoliation with tape on a PDMS stamp. The monolayer TMDs were then transferred at room temperature onto a holey SiN TEM support. HAADF- and MAADF-STEM images were acquired with an aberration-corrected JEM-ARM200F (JEOL) instrument equipped with a cold-field emission gun and an ASCOR probe corrector, set at an acceleration voltage of 80~kV. HAADF- and MAADF-STEM images were collected at semiangles of 68--280 and 30--120~mrad, respectively, with a beam convergence angle of 31~mrad and a camera length of 8~cm.

\subsection{Patch size determination}

We determine the patch size based on the autocorrelation map of the experimental image. First, we compute the 2D autocorrelation map and take its radial average (centered at the origin) to obtain a 1D profile. The first peak in this radial profile corresponds to the nearest-neighbor distance in the lattice. For hexagonal lattices, this is equal to the lattice constant $a$. To capture sufficient local structure, we set the patch size to approximately $\sqrt{3}a$.



\subsection{Training details}

We generated approximately 55 million local patches via multislice simulation and encoded each into a 31-dimensional feature vector. These were split 80/20 into training ($\approx$44 million samples) and test ($\approx$11 million samples) sets.

We implemented three attention models (attn-4k, attn-200k, and attn-400k). All models share an input dimension of 31. Each was trained for 10 epochs using the Adam optimizer with a learning rate of 0.001, a cross-entropy loss function, and a single dropout layer (rate=0.1). The models differ in batch size: 4,096 for attn-4k, 204,800 for attn-200k, and 409,600 for attn-400k.

In parallel, we trained three multilayer perceptron (MLP) variants: MLP-NoContext, MLP-Simple, and MLP-Deep. All were trained for 20 epochs with a batch size of 4,096, using the Adam optimizer (learning rate = 0.001), cross-entropy loss, and a single dropout layer (rate=0.1). They differ in input dimension: MLP-NoContext has an input dimension of 6, while MLP-Simple and MLP-Deep have an input dimension of 31.

\textbf{Author contributions}: J.D. conceived the research concept and designed the context-aware learning framework in consultation with N.D.L. J.D. and C.Z. developed the computational algorithms and performed multislice STEM simulations. L.L. synthesized the doped TMD samples and acquired STEM experimental data. G.E. and M.B. provided guidance on experimental design and interpretation. N.D.L. supervised the research. J.D. and N.D.L. wrote the manuscript with contributions from all authors.

\textbf{Code availability}: The source code and algorithms from this study are available on our public GitHub repository, \href{https://github.com/jiadongdan/motif-learn}{motif-learn}. This repository contains all the necessary scripts and tutorial notebooks needed to reproduce our experiments and apply our methods to new projects. 

\textbf{Competing interests:} The authors declare that they have no competing interests.

\textbf{Acknowledgment}: The authors thank Su Ying Quek for helpful discussions. N.D.L. acknowledges support from the Singapore Ministry of Education under Grant MOE-0071601. J.D. and N.D.L. also acknowledge funding from the 2025 NUS–AISI Joint Research Initiative Fund (A-8003325-00-00). I.V. acknowledges the support from Singapore National Research Foundation (NRF-T-CRP-2025-0007) and Agency for Science, Technology and Research (A*STAR). This project is also supported by the Eric and Wendy Schmidt AI in Science Postdoctoral Fellowship, a Schmidt Futures program.

\printbibliography

\clearpage
\setcounter{figure}{0}
\setcounter{table}{0}
\setcounter{section}{0}
\setcounter{secnumdepth}{0}
\renewcommand{\thefigure}{S\arabic{figure}}
\renewcommand{\thetable}{S\arabic{table}}
\captionsetup[figure]{font={normal,stretch=1.0},labelfont=bf}
\captionsetup[table]{font={normal,stretch=1.0},labelfont=bf}
\fancyhead[L]{\sffamily\small Supplementary Information}
\fancyfoot[R]{\sffamily\small \thepage}

\vspace*{4em}
\begin{center}
\textbf{\Large Supplementary Information}\\[2.5em]
{\LARGE\bfseries Context-Aware Deep Learning for Defect Classification\\[0.35em]
in Atomic-Resolution STEM}
\end{center}

\renewcommand{\authorvspace}{2em}

\vspace{\authorvspace}

\begin{center}
Jiadong Dan\textsuperscript{1,2,*},
Cheng Zhang\textsuperscript{3},
Leyi Loh\textsuperscript{4},
Ivan Verzhbitskiy\textsuperscript{5},
Yuan Chen\textsuperscript{6},
Goki Eda\textsuperscript{6,7},
Michel Bosman\textsuperscript{8},
N. Duane Loh\textsuperscript{1,2,6,$\dagger$}
\end{center}
\vspace{0.5em}

\vspace{-1em}
\begin{center}
\begin{minipage}{0.75\textwidth}
\small
\raggedright
\textsuperscript{1}Department of Biological Sciences, National University of Singapore, Singapore\\[0.3em]
\textsuperscript{2}Centre for BioImaging Sciences (CBIS), National University of Singapore, Singapore\\[0.3em]
\textsuperscript{3}Department of Materials Science and Engineering, City University of Hong Kong, China\\[0.3em]
\textsuperscript{4}Department of Materials Science and Metallurgy, University of Cambridge, United Kingdom\\[0.3em]
\textsuperscript{5}Quantum Innovation Centre (Q. InC), Agency for Science Technology and Research (A*STAR), Singapore\\[0.3em]
\textsuperscript{6}Department of Physics, National University of Singapore, Singapore\\[0.3em]
\textsuperscript{7}Department of Chemistry, National University of Singapore, Singapore\\[0.3em]
\textsuperscript{8}Department of Materials Sciences and Engineering, National University of Singapore, Singapore\\[0.3em]
\end{minipage}
\end{center}
\vspace{1em}

\renewcommand{\thefootnote}{}
\footnotetext{\textsuperscript{$*$, $\dagger$}Corresponding authors: jiadong.dan@u.nus.edu; duaneloh@nus.edu.sg}
\renewcommand{\thefootnote}{\arabic{footnote}}

\vfill
\clearpage

\section{S1. Formal Definition of Ill-Posedness in Context-Free Inference}

An inverse problem seeks to infer a latent variable $y$ (e.g., defect type) from an observation $x$ (e.g., image contrast). In the Bayesian framework, this inference is described by the conditional distribution
\begin{equation}
p(y\mid x)=\frac{p(x\mid y)p(y)}{p(x)}.
\end{equation}
A problem is said to be \textit{well-posed} (in the Hadamard sense) if the following three conditions hold:
\begin{enumerate}
    \item \textbf{Existence:} a solution $y$ exists for every $x$;
    \item \textbf{Uniqueness:} the solution $y$ is unique;
    \item \textbf{Stability:} small perturbations in $x$ lead to small changes in the inferred $y$.
\end{enumerate}

In many imaging modalities, including atomic-resolution STEM, the mapping $x\!\rightarrow\!y$ violates uniqueness: different physical or experimental conditions $(y_1,c_1)$ and $(y_2,c_2)$ can produce indistinguishable image contrasts $x_1\!\approx\!x_2$. In such cases, the likelihood term $p(x\mid y)$ becomes non-injective, and the posterior $p(y\mid x)$ becomes multi-modal or diffuse. Consequently, the inverse mapping is \textit{ill-posed}.

When a neural network $f_\theta(x)$ is trained to approximate this inverse relationship using supervised learning, the ill-posedness manifests as inconsistent or unstable predictions: the same or visually similar inputs $x$ can be assigned different labels $y$ depending on implicit correlations in the training data. In other words, $f_\theta$ converges to a context-specific but physically ambiguous mapping that fails to generalize across experimental variations.

To restore well-posedness, we introduce \textit{contextual conditioning} by augmenting the inference with observable variables $c$ describing the experimental or physical setup (e.g., beam energy, detector geometry, or chemical composition). The revised posterior becomes
\begin{equation}
p(y\mid x,c)=\frac{p(x\mid y,c)p(y\mid c)}{p(x\mid c)}.
\end{equation}
Since $c$ is known and fixed for each observation, conditioning effectively partitions the data space into subdomains where the mapping $(x,c)\!\rightarrow\!y$ is approximately one-to-one (injective). Correspondingly, a context-aware network $f_\theta(x,c)$ learns a more stable and physically grounded mapping that is consistent across materials and imaging conditions. Formally, conditioning reduces the entropy of the posterior,
\begin{equation}
H[y\mid x,c]\le H[y\mid x],
\end{equation}
thereby rendering the inference problem well-posed within each contextual domain.

Intuitively, contextual variables constrain the learned solution space by linking image features to the underlying physical parameters. Without them, a neural network must implicitly account for all possible imaging configurations within a single model; with them, the mapping is explicitly conditioned on the relevant physics and becomes stable, interpretable, and transferable.

It is important to note that the likelihood $p(x\mid y)$ and posterior $p(y\mid x)$ are probability distributions rather than deterministic functions. They describe statistical relationships between observations and latent variables, not strict one-to-one mappings. However, in the context of inverse problems, these distributions imply functional relationships between physical parameters and observed data. When we refer to the ``mapping'' defined by $p(x\mid y)$ or $p(y\mid x)$, we mean the underlying association between variables that these probabilistic models encode. In practice, a neural network $f_\theta(x)$ (or $f_\theta(x,c)$ when conditioned on context) learns a deterministic approximation to this relationship—typically corresponding to the posterior mean or maximum a posteriori (MAP) estimate derived from $p(y\mid x,c)$.

\section{S2. Empirical Estimation of Posterior Entropy}

The conditional entropy of the target variable $y$ given observations $x$ (and optionally context $c$) quantifies the uncertainty of the posterior distribution. 
Lower conditional entropy indicates a more confident and therefore better-posed inference. 
To compare context-free and context-aware models, we estimate these quantities empirically from the predicted posteriors.

\begin{equation}
H[y\mid x] = \mathbb{E}_{x}\!\left[-\sum_{k=1}^{K} p(y=k\mid x)\log p(y=k\mid x)\right],
\end{equation}

\begin{equation}
H[y\mid x,c] = \mathbb{E}_{x,c}\!\left[-\sum_{k=1}^{K} p(y=k\mid x,c)\log p(y=k\mid x,c)\right].
\end{equation}

Given a trained classifier that outputs softmax probabilities 
$p_\theta(y\mid x)\in\mathbb{R}^{K}$ or $p_\theta(y\mid x,c)\in\mathbb{R}^{K}$ for $N$ samples, 
the empirical estimates are computed as

\begin{equation}
\widehat{H[y\mid x]} = -\frac{1}{N}\sum_{i=1}^{N}\sum_{k=1}^{K} p_\theta(y=k\mid x_i)\log p_\theta(y=k\mid x_i),
\end{equation}

\begin{equation}
\widehat{H[y\mid x,c]} = -\frac{1}{N}\sum_{i=1}^{N}\sum_{k=1}^{K} p_\theta(y=k\mid x_i,c_i)\log p_\theta(y=k\mid x_i,c_i).
\end{equation}

All logarithms are natural, and the entropy is expressed in nats (use $\log_2$ for bits). 
The entropy reduction is defined as

\begin{equation}
\Delta H = \widehat{H[y\mid x]} - \widehat{H[y\mid x,c]},
\end{equation}

which provides an empirical measure of how much contextual information decreases posterior uncertainty.
A positive $\Delta H$ confirms that conditioning on $c$ sharpens the posterior distribution $p(y\mid x,c)$ and thus improves the well-posedness of the inference task.

\pagebreak
\section{S3. Relationship Between BIC and Log-Model Evidence}

The Bayesian Information Criterion (BIC) provides an asymptotic approximation of the \textbf{Log-Model Evidence}, $\ln \mathcal{P}(\mathbf{D} \mid \mathcal{M})$, which is the probabilistic foundation for Bayesian model selection.

The general BIC formula is:
\[
\text{BIC} \approx -2 \ln \mathcal{L}(\hat{\boldsymbol{\theta}}) + k \ln(N)
\]
where $\ln \mathcal{L}(\hat{\boldsymbol{\theta}})$ is the \textbf{total maximized log-likelihood} across all $N$ data points, $k$ is the number of parameters, and $N$ is the sample size.

\subsection*{The $N$-Dependence in the Fit Term}

The total maximized log-likelihood, $\ln \mathcal{L}(\hat{\boldsymbol{\theta}})$, is the sum of the log-likelihoods for each sample, thus inheriting a linear dependence on $N$:
\[
\ln \mathcal{L}(\hat{\boldsymbol{\theta}}) = \sum_{i=1}^{N} \ln \mathcal{P}(\mathbf{y}_i \mid \mathbf{x}_i, \hat{\boldsymbol{\theta}}) \approx -N \cdot \mathcal{L}_{CE}^{\text{min}}
\]
where $\mathcal{L}_{CE}^{\text{min}}$ is the minimum average cross-entropy loss (average negative log-likelihood per sample).

Substituting this relationship into the BIC formula makes the $N$-dependence explicit for practical calculation:
\[
\text{BIC} \approx \underbrace{2N \cdot \mathcal{L}_{CE}^{\text{min}}}_{\substack{\text{Fit Term:} \\ \text{Linear dependence on } N}} + \underbrace{k \ln(N)}_{\substack{\text{Complexity Term:} \\ \text{Logarithmic dependence on } N}}
\]
This structure clearly shows that a large $N$ ensures that the quality of the model's fit ($\mathcal{L}_{CE}^{\text{min}}$) is the dominant factor in the BIC score, establishing the relationship:
\[
\text{BIC} \approx -2 \ln \mathcal{P}(\mathbf{D} \mid \mathcal{M}) + \text{Constant}
\]
Consequently, a \textbf{lower BIC score} corresponds to a \textbf{higher Log-Model Evidence}.

\begin{figure}[p]
\centering
\includegraphics[width=0.7\textwidth]{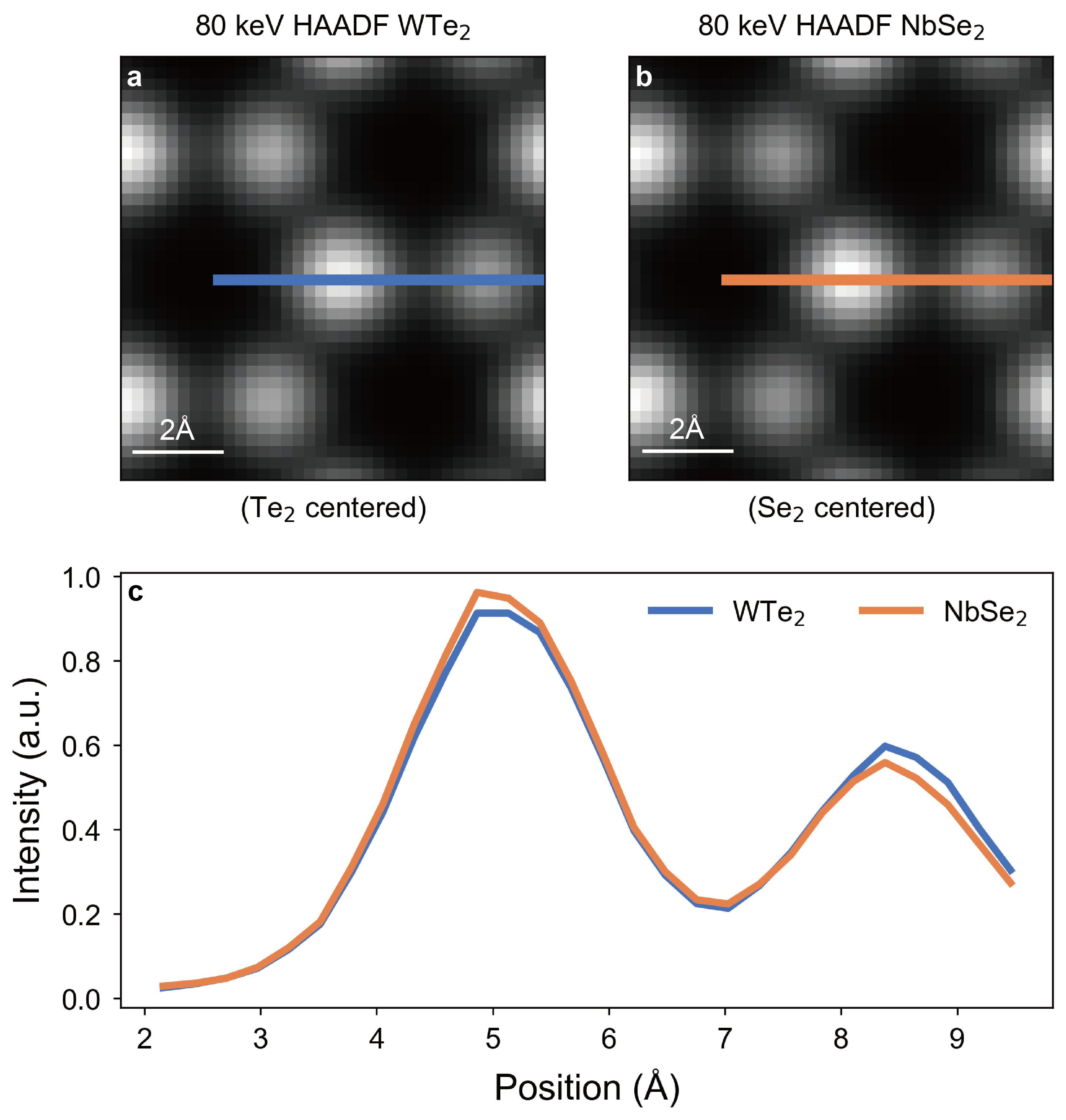}
\caption{\textbf{Simulated HAADF-STEM images of WTe\textsubscript{2} and NbSe\textsubscript{2} at 80 keV, illustrating similar atomic contrast.} (\textbf{a}) Simulated image of WTe\textsubscript{2}, centered on a Te\textsubscript{2} column. (\textbf{b}) Simulated image of NbSe\textsubscript{2}, centered on a Se\textsubscript{2} column. (\textbf{c}) Line intensity profiles along the blue and orange lines in (a) and (b), respectively, highlight the similarity in contrast between the two materials.}\label{SI_similar_contrast}
\end{figure}

\begin{figure}[p]
\centering
\includegraphics[width=0.8\textwidth]{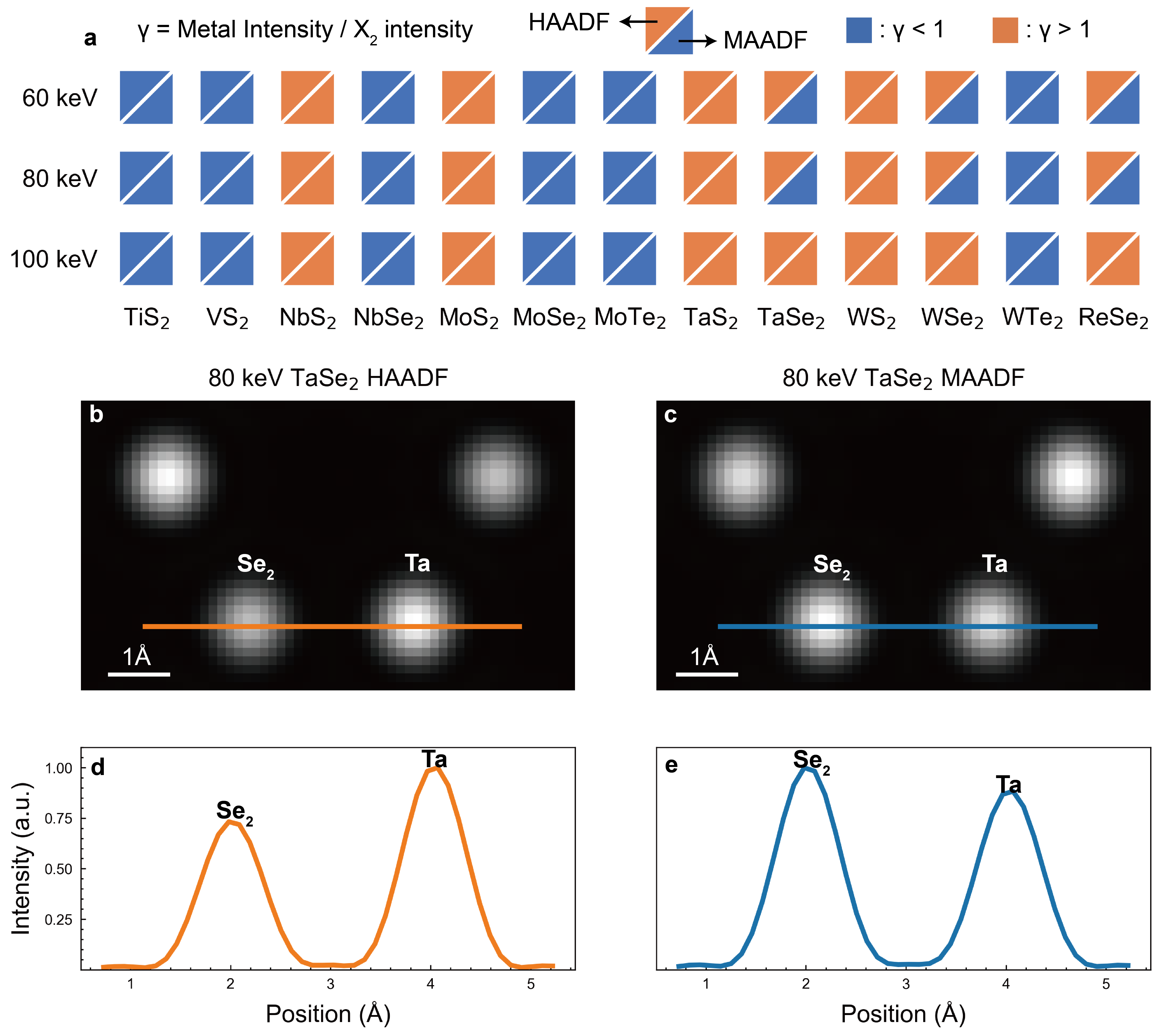}
\caption{\textbf{Metal to X\textsubscript{2} atoms contrast ratio reverse under different detectors.} (\textbf{a}) Intensity ratio ($\gamma$) of transition metal M and dichalcogen X\textsubscript{2} at 60 keV, 80 keV and 100 keV. Top left triangle indicates HAADF mode, lower right triangle indicate MAADF mode. Blue color means ratio less than one, orange color means ratio larger than one. (\textbf{b}) Simulated HAADF image of TaSe\textsubscript{2} at 80 keV. (\textbf{c}) Simulated MAADF image of TaSe\textsubscript{2} at 80 keV. (\textbf{d}) The line profile shows the intensities of Se\textsubscript{2} and Ta atomic columns in panel b. (\textbf{e}) The line profile shows the intensities of Se\textsubscript{2} and Ta atomic columns in panel c.}\label{SI_contrast_reversal}
\end{figure}

\begin{figure}[p]
\centering
\includegraphics[width=0.8\textwidth]{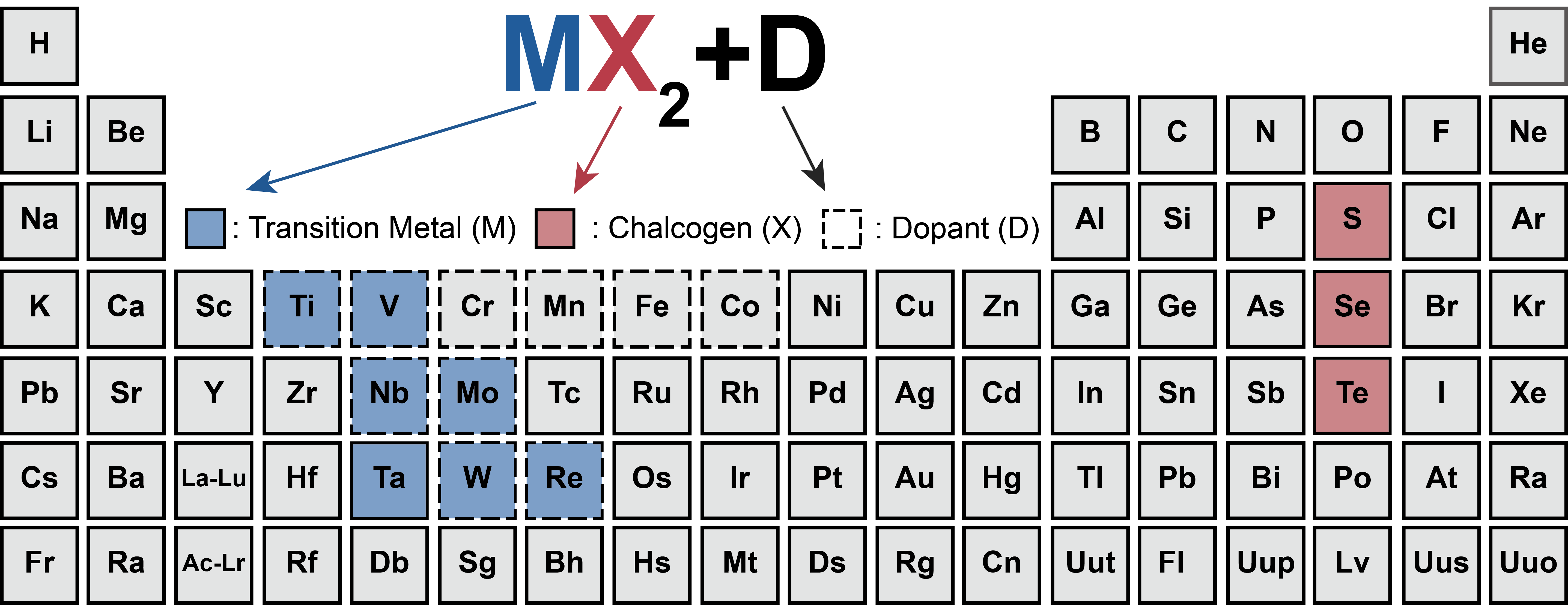}
\caption{\textbf{Overview of experimentally observed 1H-phase transition metal dichalcogenides (TMDs) in this study.} We considered all 13 $\times$ 10 = 130 possible MX\textsubscript{2} + D pairings (M = transition metal, X = chalcogen, D = dopant element). Pairings in which D and M lie in the same period of the periodic table were omitted (marked ``$\times$''), leaving 96 valid combinations. These are shown as purple cells.}\label{SI_elementary_table}
\end{figure}

\begin{figure}[p]
\centering
\includegraphics[width=0.8\textwidth]{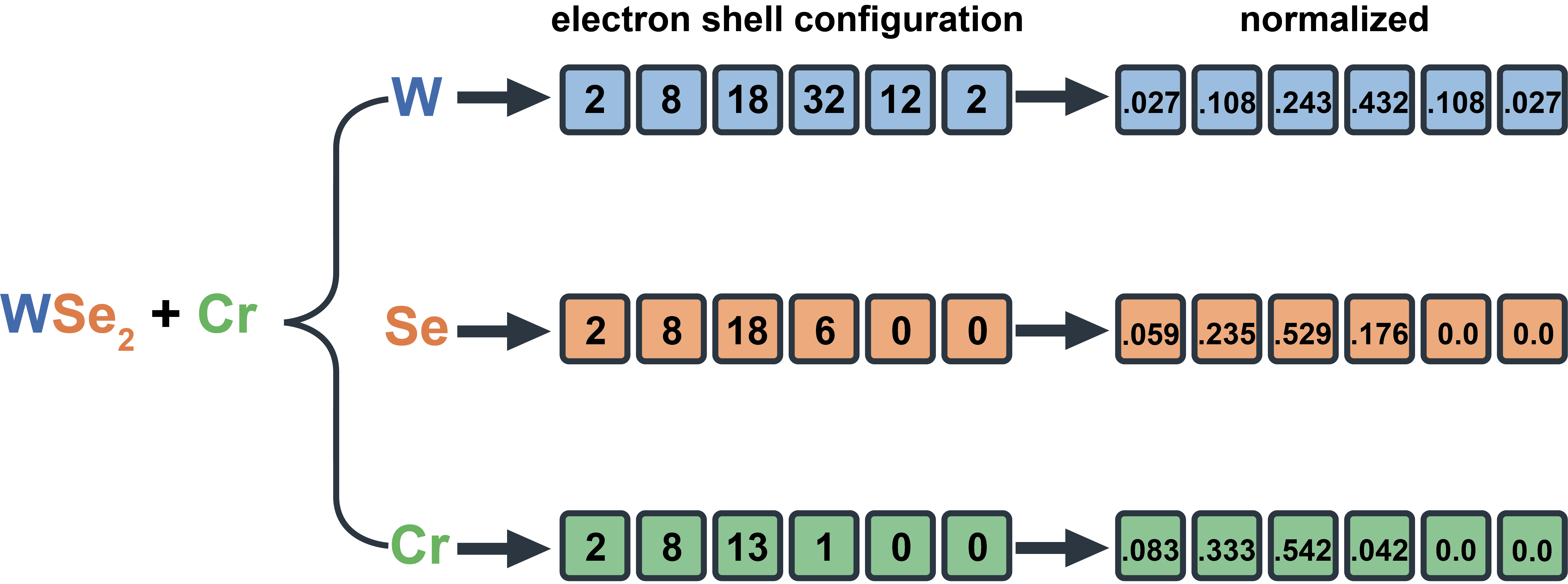}
\caption{\textbf{Encoding chemical information into a fixed-length array using electron shell configurations.} We represent each MX\textsubscript{2}+D compound as an 18-element vector: six values for the transition metal (M), six for the chalcogen (X), and six for the dopant (D). Each block of six entries records the electron shell configuration, normalized by the atom's atomic number Z. For instance, tungsten (Z = 74) has shell populations \{2, 8, 18, 32, 18, 8\}, which become \{0.027, 0.108, 0.243, 0.432, 0.243, 0.108\} after division by 74 (these sum to 1).}\label{SI_chemical_encoding}
\end{figure}

\begin{figure}[p]
\centering
\includegraphics[width=0.8\textwidth]{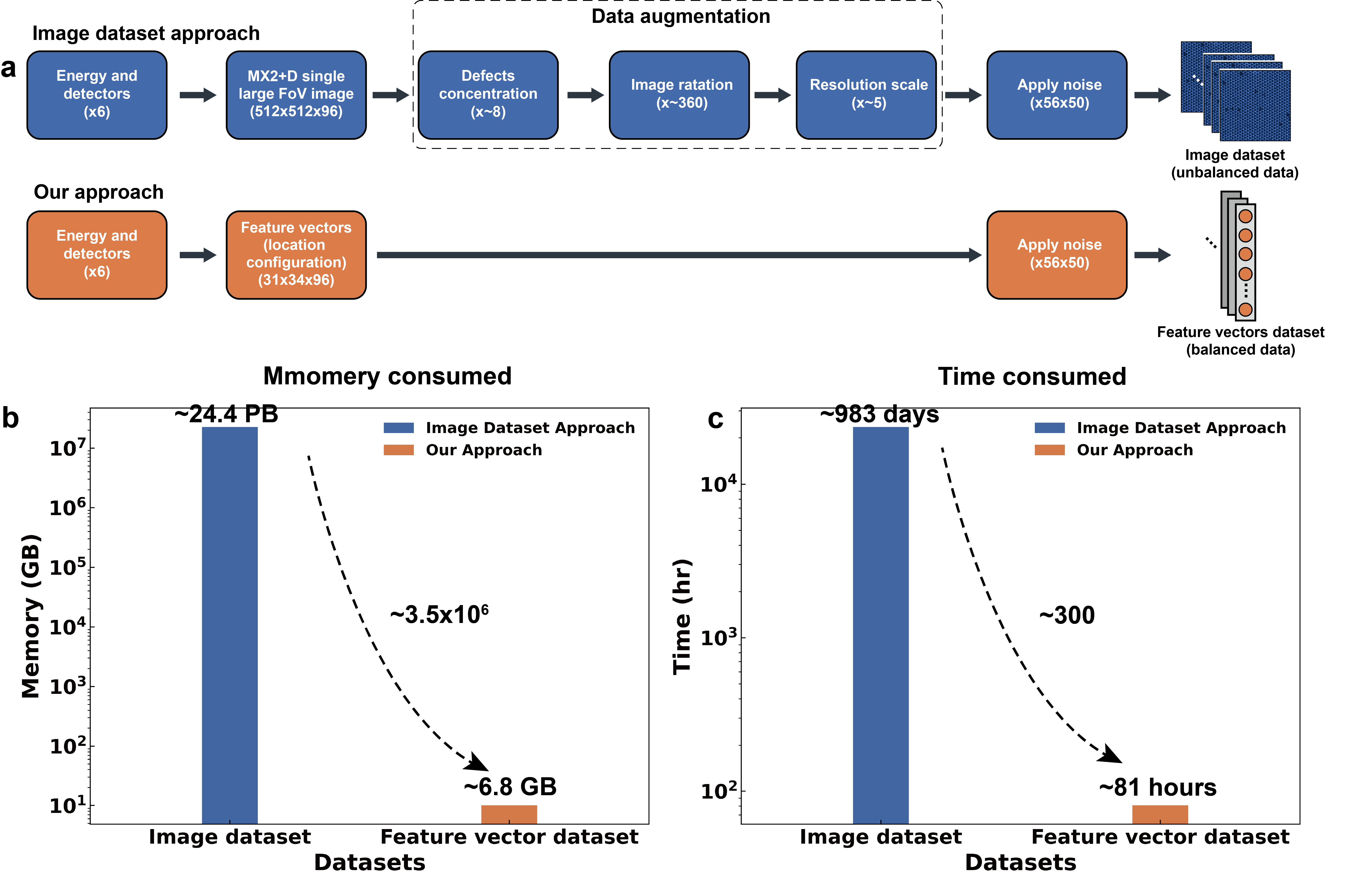}
\caption{\textbf{Comparison of dataset size and multislice simulation cost.} The proposed method requires fewer images and substantially less CPU time than the conventional image-based approach.}\label{SI_datasize_comparison}
\end{figure}

\begin{figure}[p]
\centering
\includegraphics[width=0.8\textwidth]{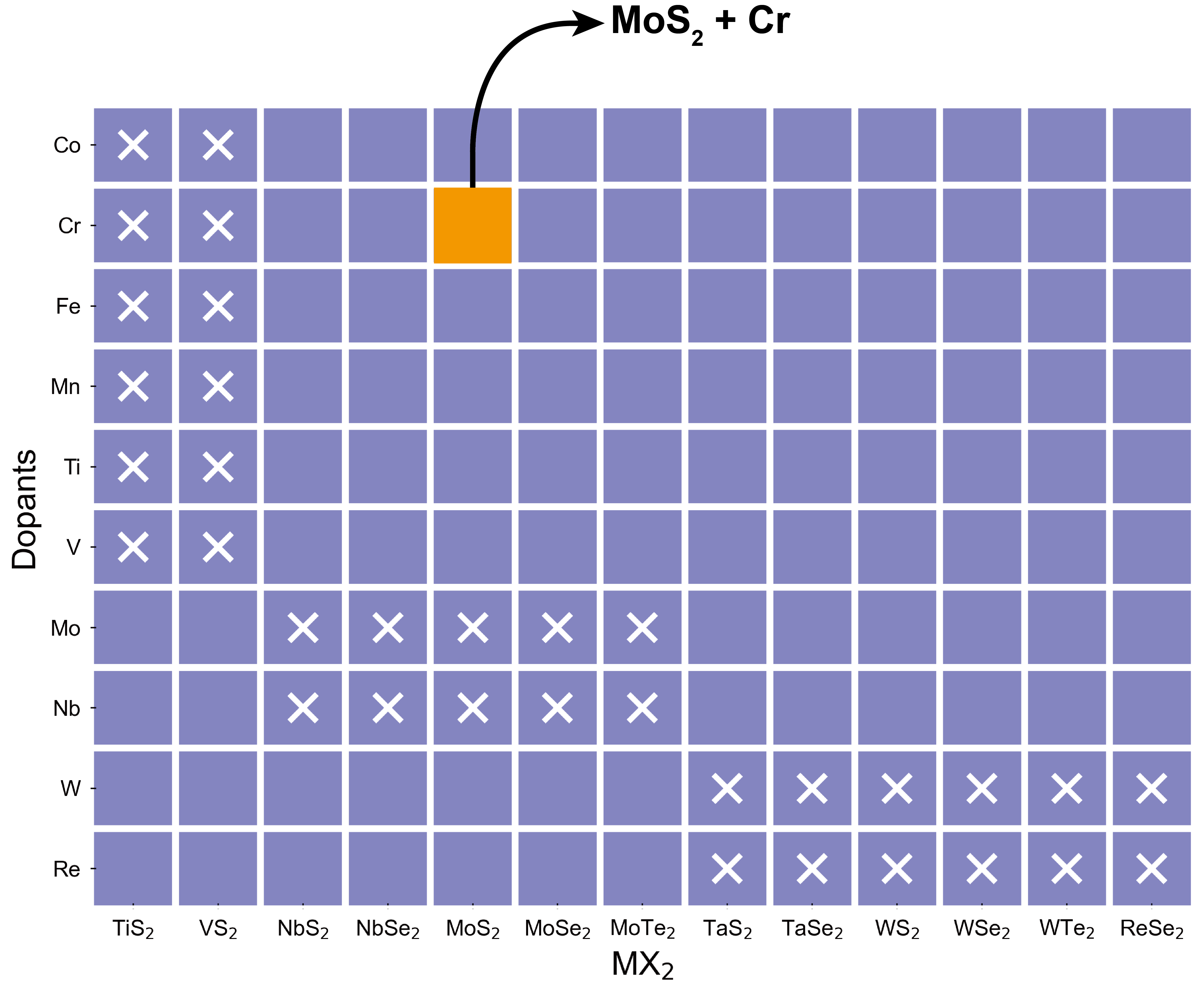}
\caption{\textbf{Overview of 96 doped cases of doped 1H-phase transition-metal dichalcogenides (MX\textsubscript{2} + D).} The periodic table highlights the elements involved: transition metals (M = Ti, V, Cr, Nb, Mo, Ta, W) are shaded in solid blue, chalcogens (X = S, Se, Te) in solid red, and ten dopant elements (D) are indicated by black dashed borders around their element boxes.}\label{SI_MX+D_combination}
\end{figure}

\begin{figure}[p]
\centering
\includegraphics[width=0.8\textwidth]{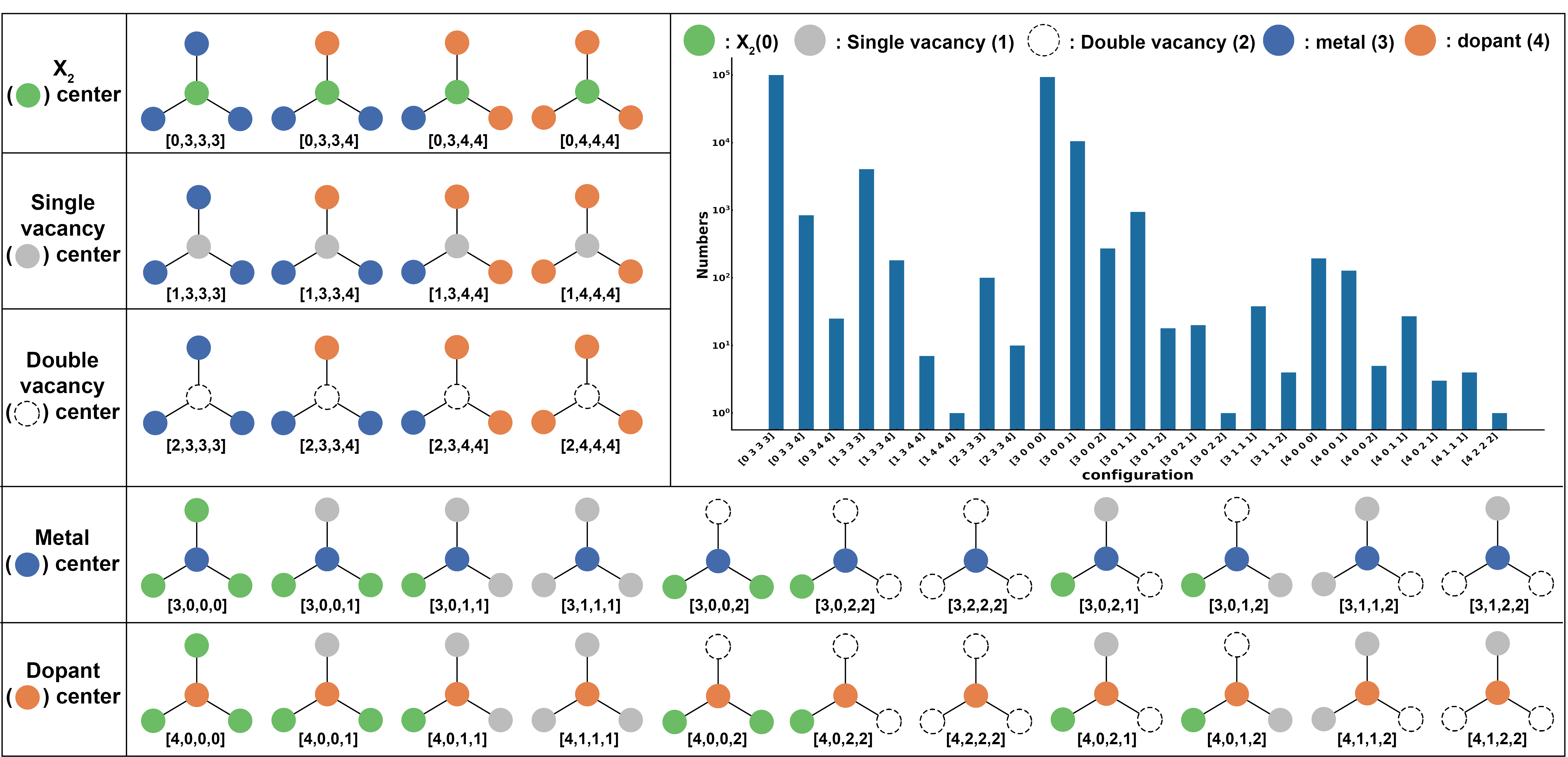}
\caption{\textbf{Thirty-four structural configurations and their occurrences in the experimental dataset.}}\label{SI_motif_configuration}
\end{figure}

\begin{figure}[p]
\centering
\includegraphics[width=0.8\textwidth]{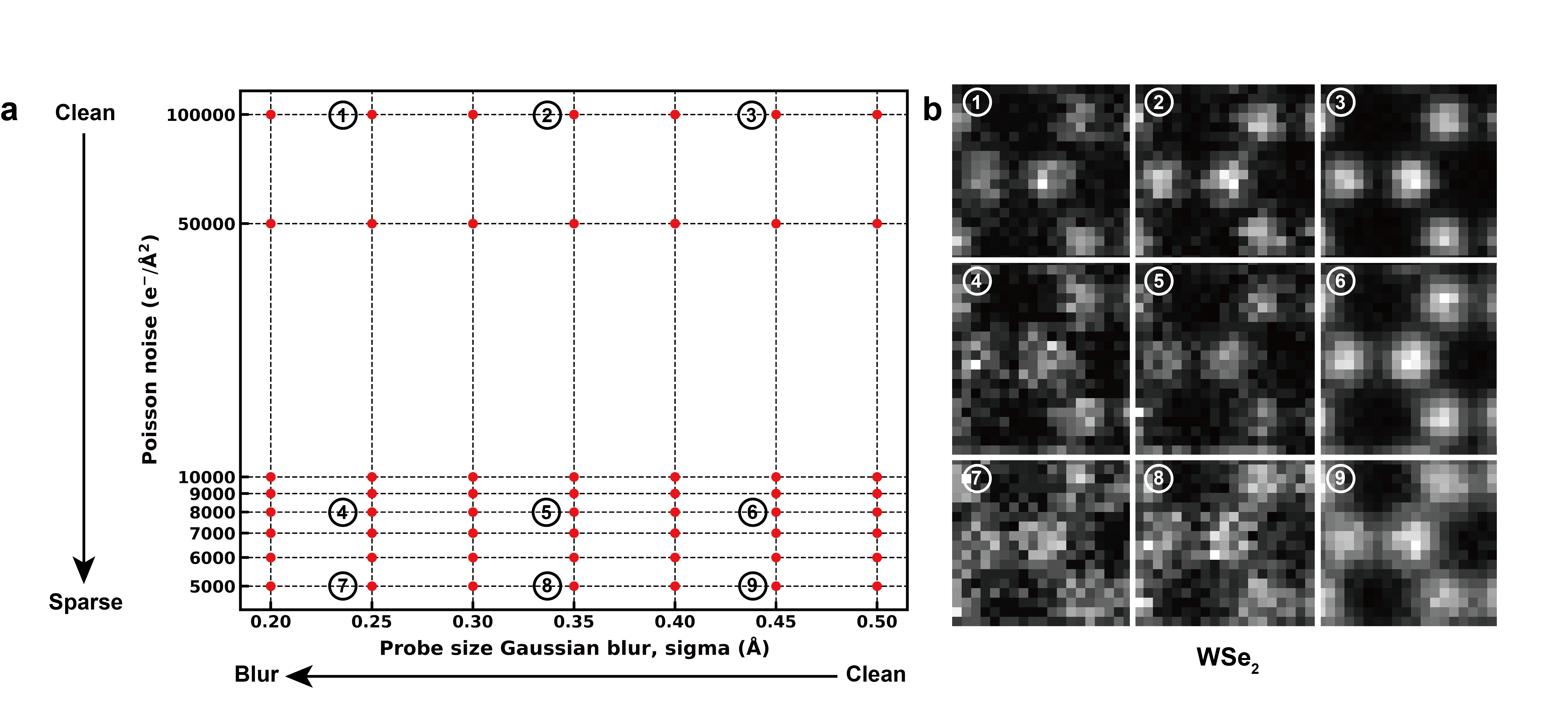}
\caption{\textbf{Summary of noise and augmentation parameters applied in STEM simulations.} Gaussian blur scales with probe size, while Poisson noise parameters are varied to reproduce realistic experimental conditions.}\label{SI_grid_noise}
\end{figure}

\begin{figure}[p]
\centering
\includegraphics[width=0.8\textwidth]{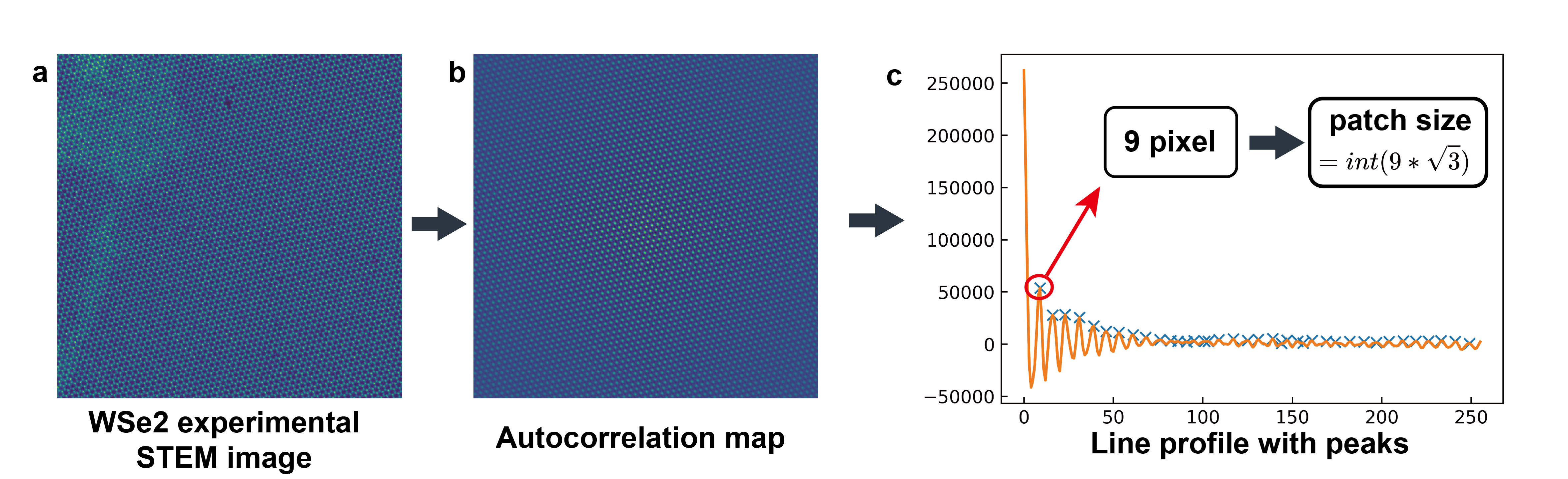}
\caption{\textbf{The workflow for selecting appropriate patch size.}}\label{SI_exp_data_worlflow}
\end{figure}

\begin{figure}[p]
\centering
\includegraphics[width=0.8\textwidth]{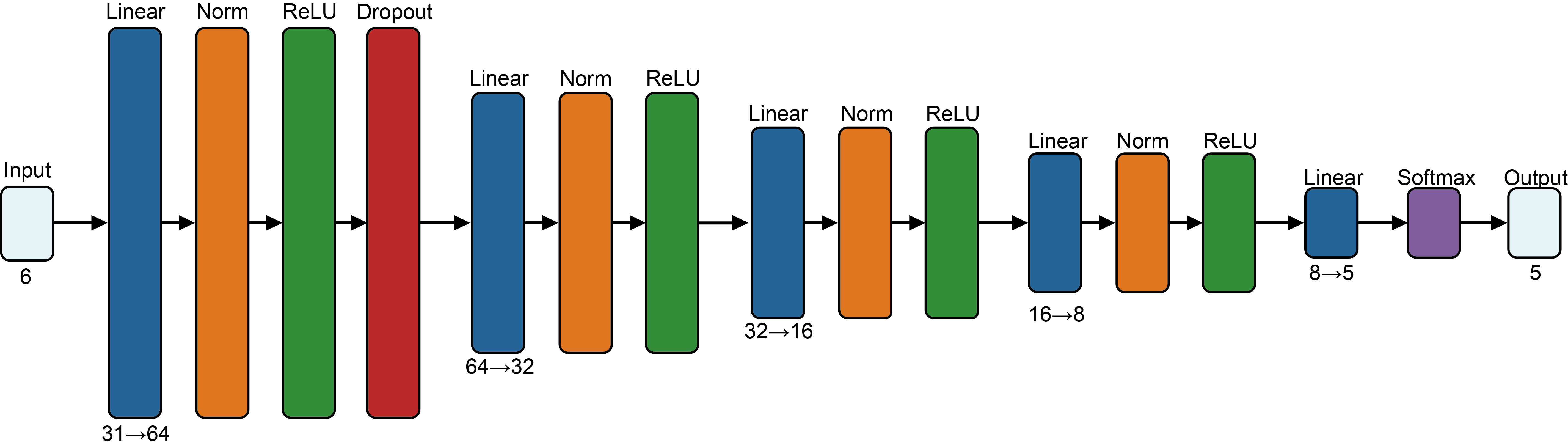}
\caption{\textbf{\texttt{MLP-NoContext} model architecture.} The input features have a dimension of 6, which do not contain context information.}\label{SI_MLP_No_Context_model}
\end{figure}

\begin{figure}[p]
\centering
\includegraphics[width=0.8\textwidth]{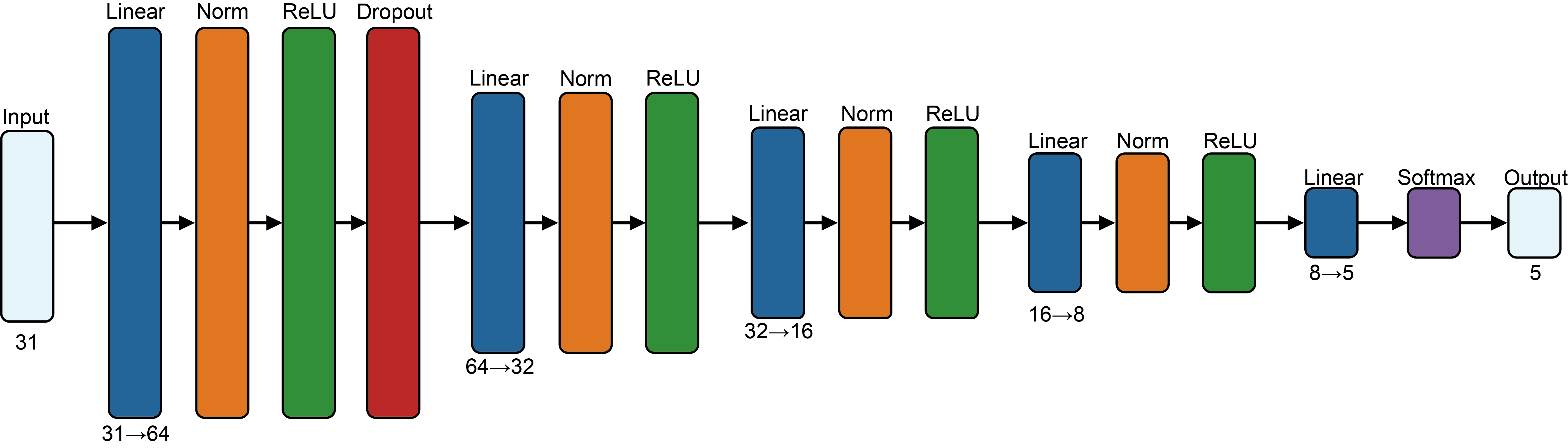}
\caption{\textbf{Architecture of the \texttt{MLP-Simple} model.} The input feature vector has a dimension of 31.}
\label{MLP_simple}
\end{figure}

\begin{figure}[p]
\centering
\includegraphics[width=0.8\textwidth]{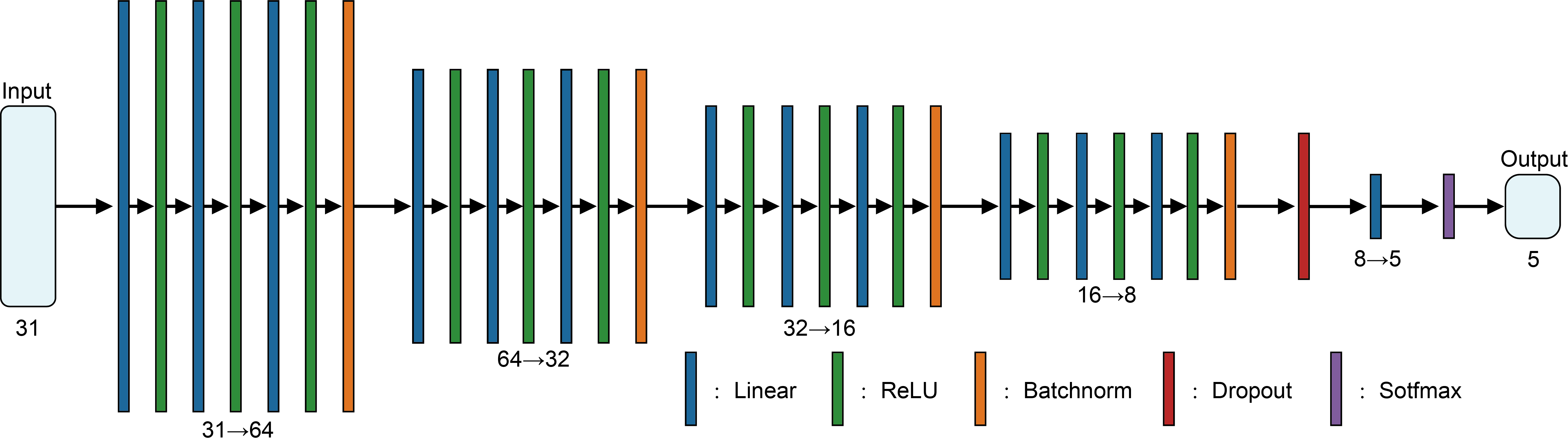}
\caption{\textbf{Architecture of the \texttt{MLP-Deep} model.} The input feature vector has a dimension of 31.}
\label{MLP_deep}
\end{figure}

\begin{figure}[p]
\centering
\includegraphics[width=0.8\textwidth]{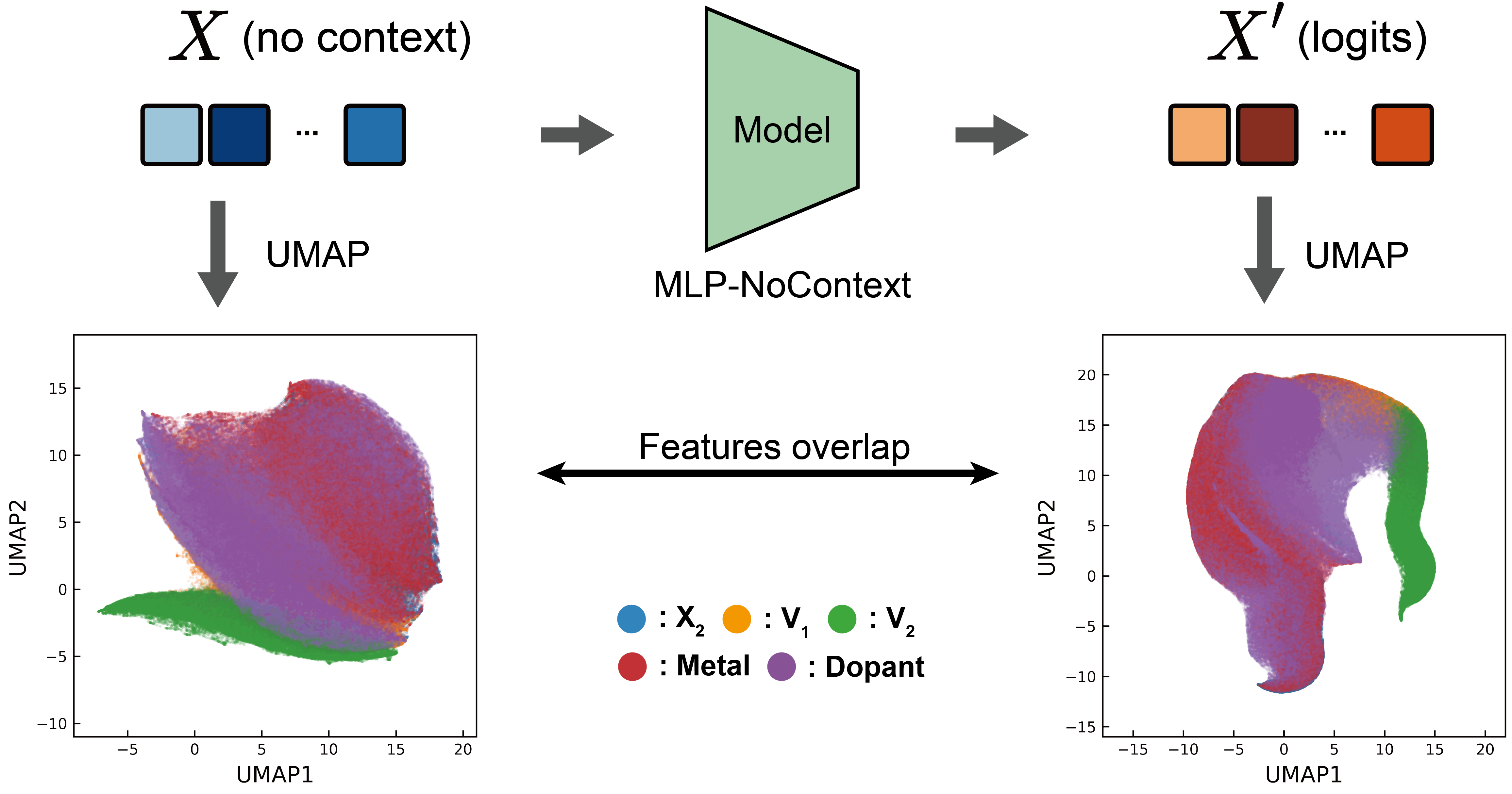}
\caption{\textbf{UMAP projections of feature distributions for input data without context information and logits after processing through the \texttt{MLP-NoContext} model.} Both UMAP figures show the overlapping among all five classes.}\label{SI_MLP_No_Context_umap}
\end{figure}

\begin{figure}[p]
\centering
\includegraphics[width=0.8\textwidth]{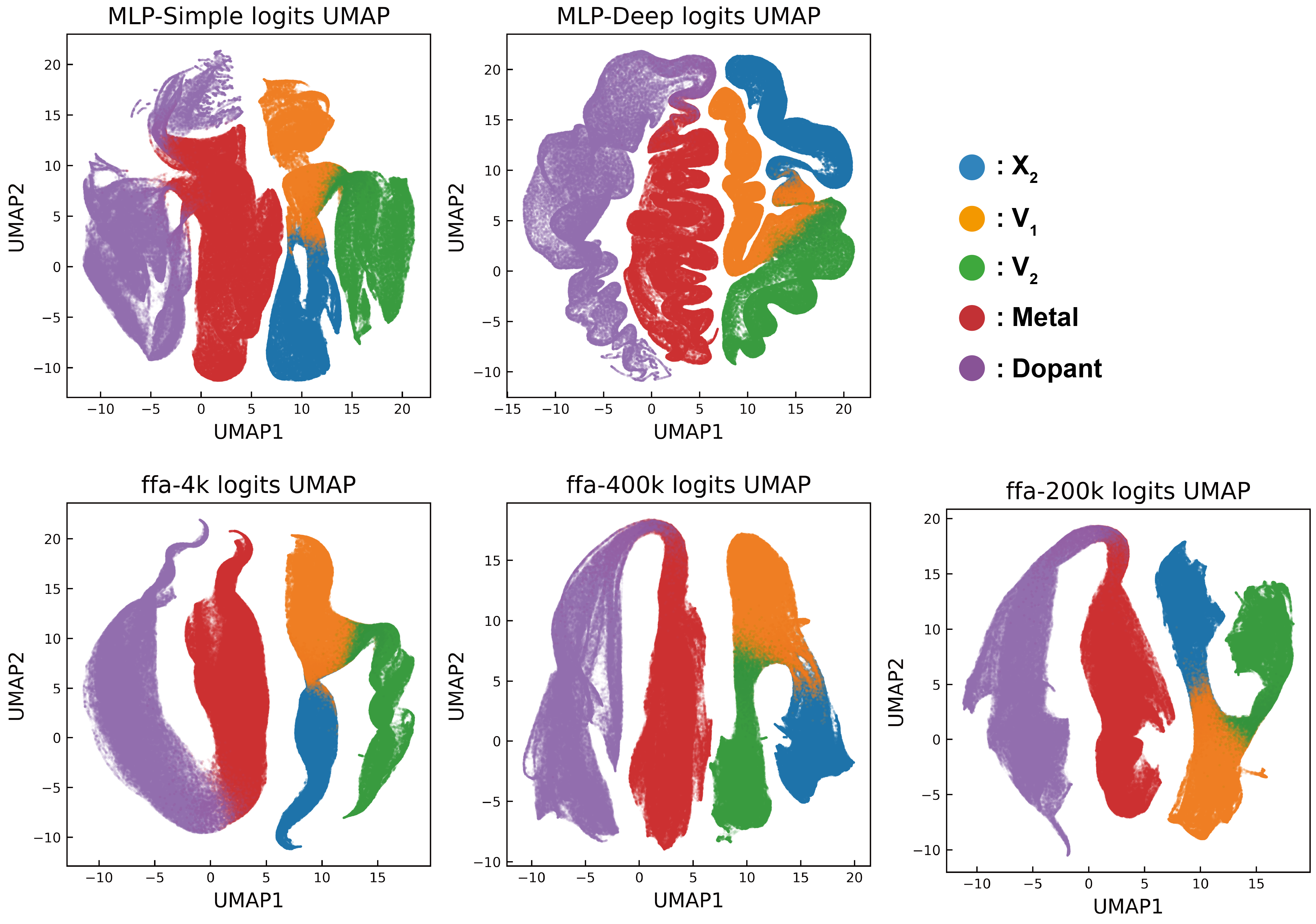}
\caption{\textbf{UMAP projections of logits for data with context information processed by five models, showing clear separation of all five classes into distinct clusters.}}
\label{logits_UMAP}
\end{figure}

\begin{figure}[p]
\centering
\includegraphics[width=0.8\textwidth]{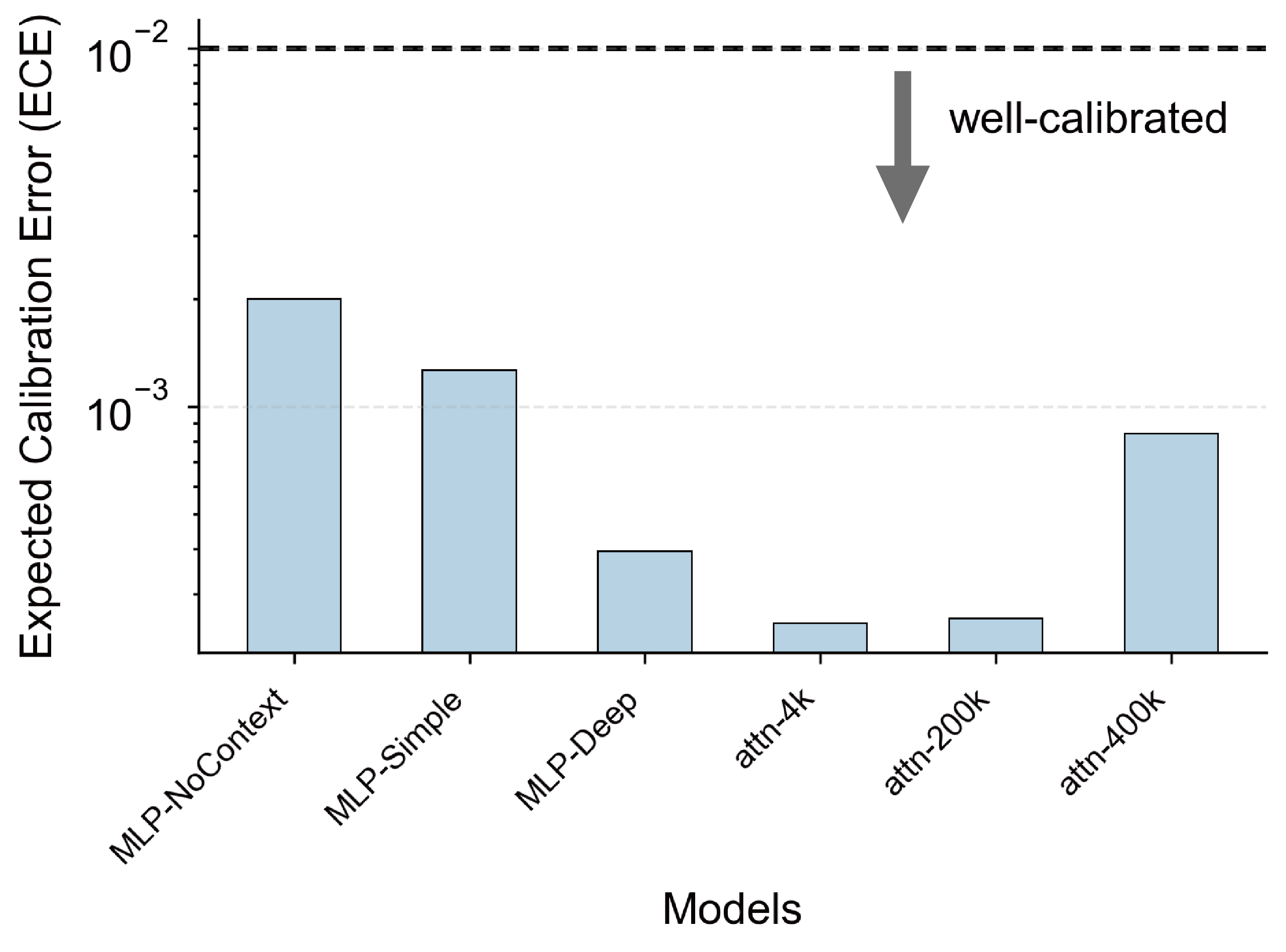}
\caption{\textbf{Bar chart of Expected Calibration Error (ECE) scores for six models. All six models are well-calibrated.}}
\label{ECE_bar}
\end{figure}

\begin{figure}[p]
\centering
\includegraphics[width=0.8\textwidth]{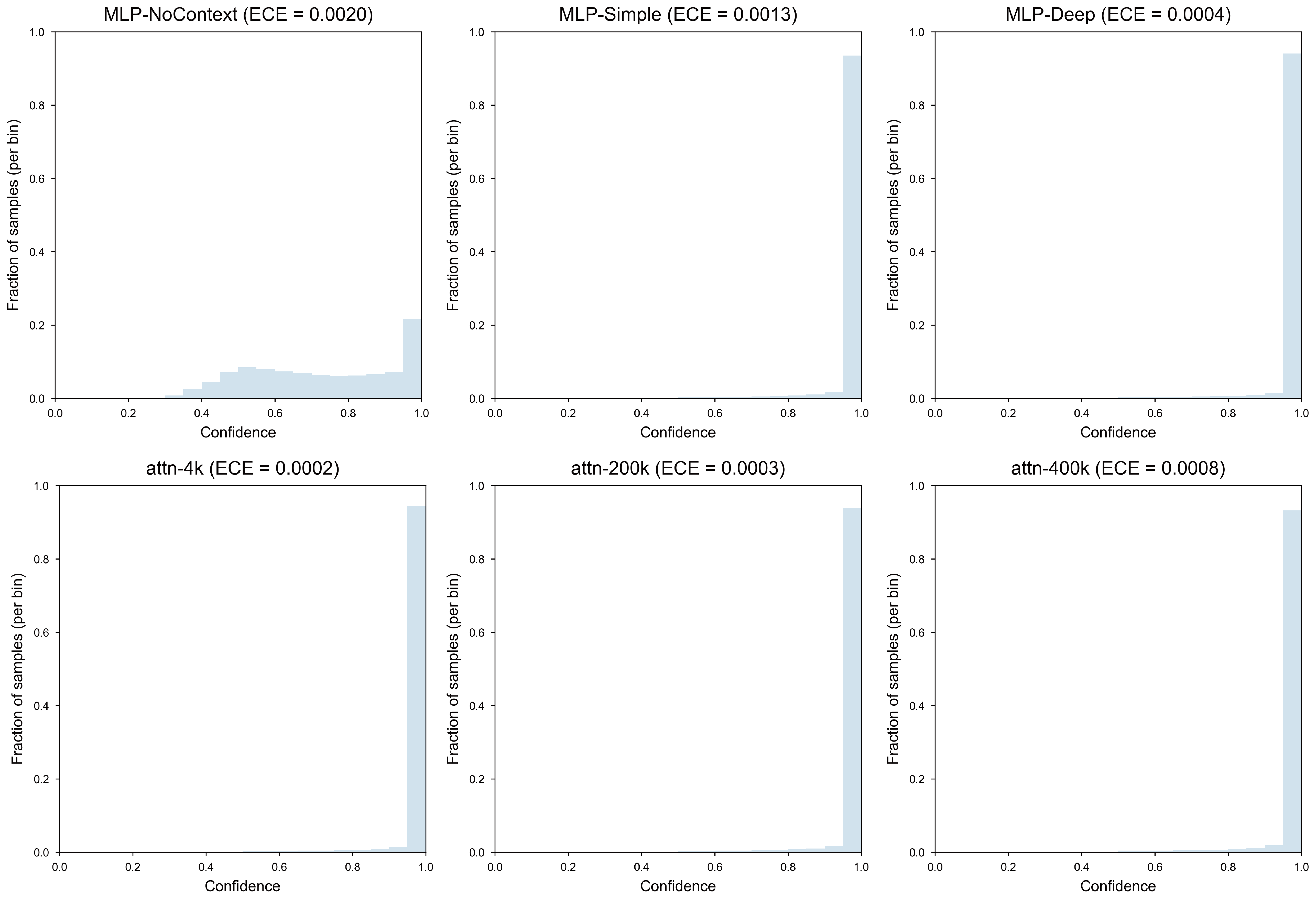}
\caption{\textbf{The Confidence histograms of six models.} The \texttt{MLP-NoContext} model has lower confidence than other five models.}
\label{ECE_hist}
\end{figure}

\begin{figure}[p]
\centering
\includegraphics[width=0.8\textwidth]{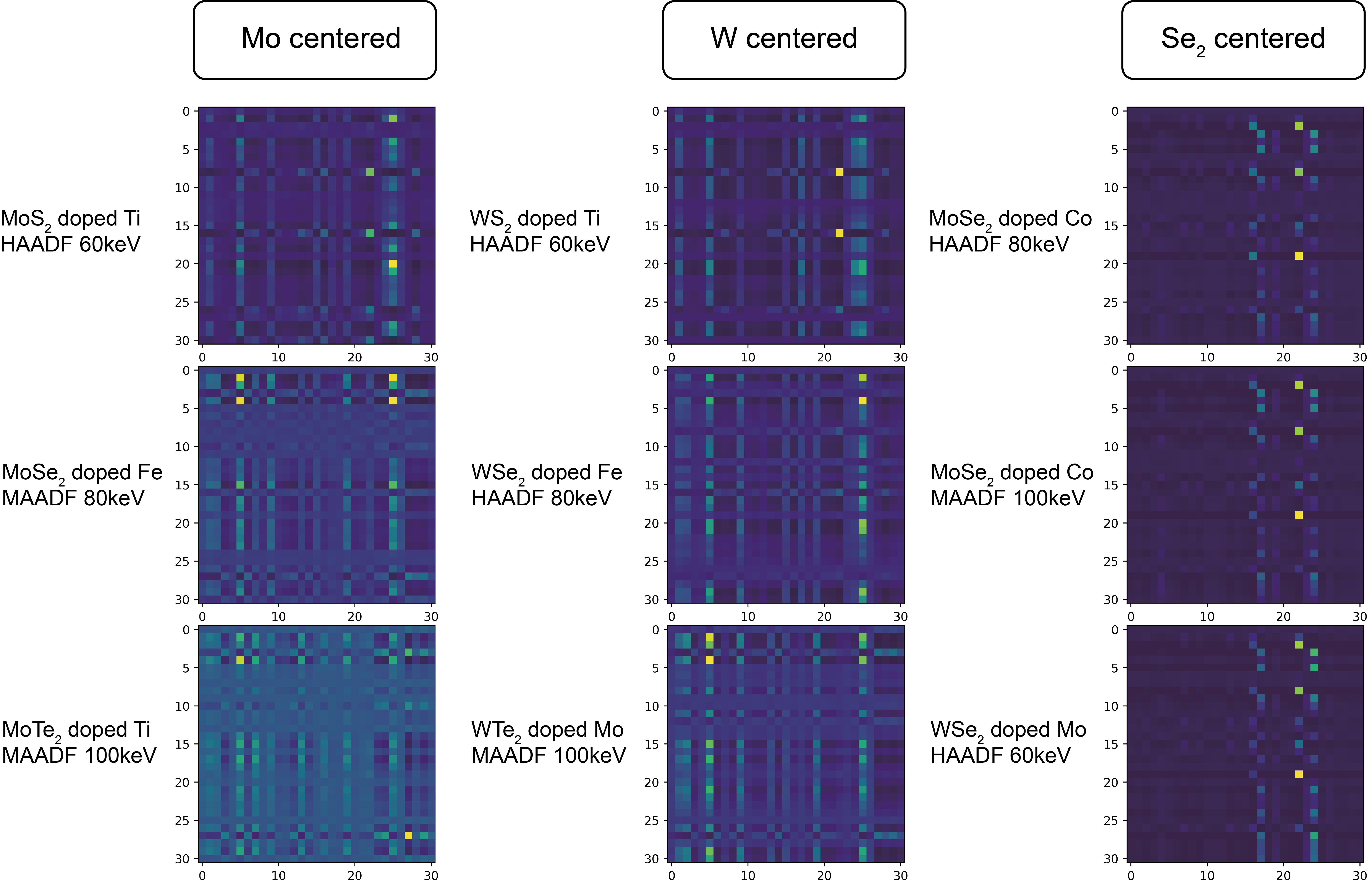}
\caption{\textbf{Third-layer attention maps of the \texttt{attn-400k} model exhibit high similarity across three conditions for each atom center.}}
\label{attn_map}
\end{figure}

\begin{figure}[p]
\centering
\includegraphics[width=0.8\textwidth]{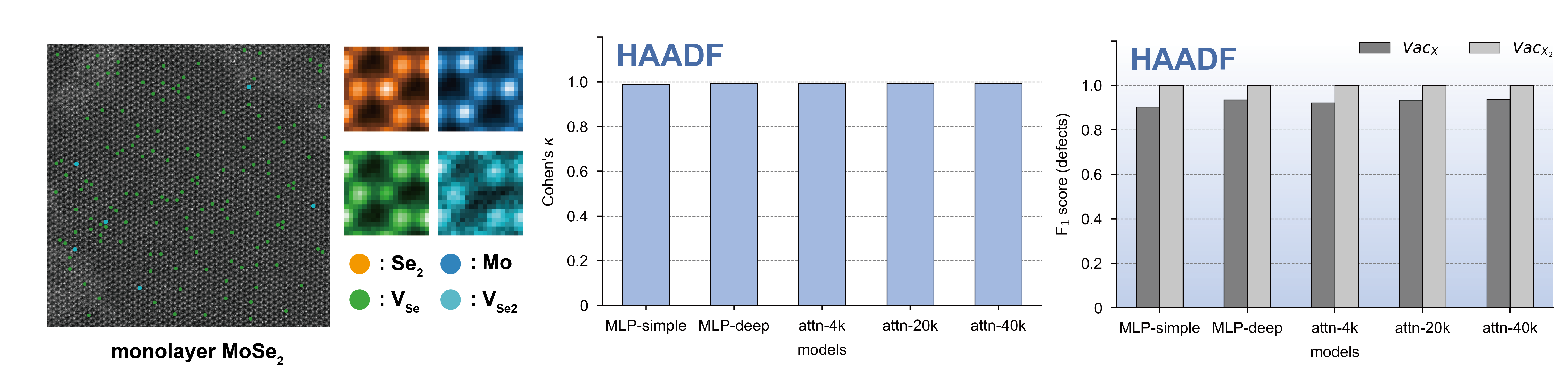}
\caption{\textbf{Representative of HAADF monolayer MoSe\textsubscript{2} image and evaluation of human and model classifications.} Monolayer MoSe\textsubscript{2} HAADF image with attn-400k model labeled defects: single Se vacancies (green), and double Se vacancies (cyan). Mean image patches for the four atomic classes--Mo sites, Se sites, single Se vacancy sites, and double Se vacancy sites--averaged over all occurrences in the example image from left image. Cohen's $\kappa$ scores, shown in the middle bar chart, quantify agreement between human consensus labels and predictions from five model variants. HAADF F\textsubscript{1} scores, in the right bar chart, illustrates two defect classes across all models.}
\label{MoSe2}
\end{figure}

\end{document}